\documentclass[lettersize,journal]{IEEEtran}
\usepackage{amsmath,amsfonts}
\usepackage{amssymb}
\usepackage{amsthm}
\usepackage{bm}

\usepackage{algorithmic}
\usepackage{color}
\usepackage{algorithm}
\usepackage{hyperref}
\usepackage{array}
\usepackage[caption=false,font=normalsize,labelfont=sf,textfont=sf]{subfig}
\usepackage{textcomp}
\usepackage{stfloats}
\usepackage{url}
\usepackage{verbatim}
\usepackage{graphicx}
\usepackage{cite}
\begin{document}


\title{Physics-Informed Uncertainty-Aware Beamforming for HAPS Massive MIMO under Imperfect CSI}

\author{{Akram~Y.~Sarhan,~\IEEEmembership{Member,~IEEE}, Osamah~A.~Abdullah,~\IEEEmembership{Member,~IEEE},\\
Khalid~T.~Musri,~\IEEEmembership{Member,~IEEE}, and Hayder Al-Hraishawi,~\IEEEmembership{Senior Member,~IEEE} \vspace{-5mm}
\thanks{
A. Y. Sarhan is with the Department of Information Technology, College of Computing and Information Technology, University of Jeddah, Saudi Arabia.\\
O. A. Abullah is with the Ministry of Higher Education and Scientific Research, Baghdad, Iraq.\\
K. T. Musri is with Department of Cybersecurity, University of Jeddah, Saudi Arabia.\\
H. Al-Hraishawi is with the Department of Electrical and Computer Engineering, University of South Florida, Tampa, FL 33620 USA.\\
Corresponding author: \emph{Akram Y. Sarhan (asarhan@uj.edu.sa)}.}
\thanks{This work was funded by the University of Jeddah, Jeddah, Saudi Arabia, under grant No. (UJ-25-DR-335). Therefore, the authors thank the University of Jeddah for its technical and financial support.}}}

\maketitle

\begin{abstract}
High-altitude platform station (HAPS) massive multiple-input multiple-output (MIMO) systems are expected to support wide-area, low-latency, and energy-efficient connectivity in future non-terrestrial networks. However, Doppler-induced channel aging, finite-rate feedback quantization, packet loss, and estimation noise impair transmitter-side channel state information (CSI), making robust downlink beamforming challenging. In HAPS channels, these impairments are strongly structured by elevation-dependent Rician propagation and line-of-sight (LoS)-dominant geometry, whereas conventional robust beamforming methods often rely on generic uncertainty models and computationally intensive optimization. This paper develops a physics-informed uncertainty-aware beamforming framework for HAPS massive MIMO systems under imperfect CSI. First, a geometry-aware channel and feedback-impairment model is developed, where CSI errors due to aging, quantization, packet loss, and noise are represented through tangent-space ellipsoidal uncertainty sets. Second, a physics-informed variational autoencoder (VAE) exploits the LoS-dominant steering manifold to enhance channel direction information and propagate learned uncertainty through unit-sphere projection. Third, the learned uncertainty representation is embedded into a robust energy-efficiency maximization formulation with probabilistic QoS awareness. To enable scalable online operation, the resulting beamforming policy is approximated using a multi-agent deterministic policy gradient framework with centralized training, decentralized execution, and differentiable power projection. Simulation results show that the proposed framework improves energy efficiency, SINR robustness, outage reliability, convergence behavior, and online runtime compared with imperfect-CSI, SDR-based, and no-VAE baselines.
\end{abstract}


\section{Introduction}
The evolution to sixth-generation (6G) wireless networks involves the integration of terrestrial and non-terrestrial infrastructures to ensure ubiquitous, resilient, and energy-efficient connectivity \cite{ saad2019vision, zhang20196g}. High-altitude platform stations (HAPS) functioning in the stratosphere at altitudes ranging from 17 to 30 km, are among the most promising enablers of this vision \cite{Toka2024}. In contrast to low Earth orbit (LEO) satellites, HAPS deliver reduced latency, adaptable repositioning, and focused regional coverage, while also providing larger footprints than terrestrial base stations \cite{Lou2023,NGSO2023}. Recent standardization efforts by 3GPP and ITU have incorporated HAPS into non-terrestrial network (NTN) frameworks, highlighting their emerging role in beyond-5G and future 6G wireless systems \cite{3gpp38811, itur_m2514}.

To exploit the wide-area coverage capability of HAPS, massive MIMO becomes a natural enabling technology rather than merely a capacity-enhancing feature, as large antenna arrays provide the beamforming gain and spatial degrees of freedom needed to support heterogeneous users over long stratospheric links \cite{Abbasi2024}. By forming narrow beams, a HAPS can compensate for severe path loss, spatially separate users with distinct spatial directions, and reduce unnecessary radiated power \cite{Shafie2024}. However, unlike terrestrial massive MIMO systems, HAPS channels are geometry-dependent and typically exhibit LoS-dominant Rician fading, with interference governed by the angular separation among users \cite{Jang2026}. Therefore, energy-efficient power control in HAPS massive MIMO systems cannot be directly adopted from terrestrial designs and must account for altitude-dependent propagation, beam-domain interference, and heterogeneous QoS requirements.

Due to the large HAPS coverage footprint, users experience substantial variations in slant range and viewing angle, resulting in highly non-uniform path loss and spatially diverse SINR conditions across the service area \cite{Jang2025}. Moreover, HAPS channels are typically LoS-dominant and are modeled using Rician fading with elevation-dependent (K)-factors \cite{Lin2026}. Users near the nadir benefit from stronger deterministic components and lower propagation loss, whereas edge users experience weaker LoS dominance, larger attenuation, and increased atmospheric impairments \cite{Karaman2025}. These spatial disparities directly influence beamforming effectiveness, inter-user interference, and the resulting power-allocation strategy.

Moreover, mobility-induced temporal dynamics impose stringent real-time constraints on HAPS downlink transmission. At high carrier frequencies, even moderate UE mobility can produce non-negligible Doppler shifts and channel-aging effects, reducing the accuracy of transmitter-side  channel state information (CSI) \cite{Pawase2026}. These effects are further exacerbated by the large HAPS coverage footprint and the associated CSI acquisition and feedback overhead. Consequently, beamforming feedback information (BFI) and channel direction information (CDI) may become outdated due to latency, finite-rate quantization, and feedback imperfections \cite{Liu2024}. Unlike terrestrial systems with dense infrastructure and frequent CSI updates, HAPS deployments must account for statistically structured CSI uncertainty in beamforming and power-control design.

At the same time, energy efficiency is a critical design requirement for HAPS platforms due to stringent onboard power and payload constraints \cite{Farooqi2025}. Although massive MIMO can improve energy efficiency through directional beamforming and spatial interference suppression, jointly optimizing transmit power, beamforming, and user reliability remains highly challenging under geometry-dependent propagation and imperfect CSI \cite{Maki2025}. Thus, energy-efficient HAPS transmission design leads to nonconvex optimization problems that are difficult to solve using conventional approaches \cite{Alqasir2024}.

\subsection{Related Work and Research Gap}
Extensive research has investigated beamforming and resource allocation for terrestrial massive MIMO systems under imperfect CSI. Classical optimization-based approaches rely on convex optimization, fractional programming, and successive convex approximation techniques to obtain suboptimal yet computationally efficient solutions for transmit power allocation and interference management \cite{Dai2025,Zhang2026,Ji2021ee,Liu2024,Kanani2025}. Although these methods provide useful performance benchmarks and theoretical insights, their computational complexity grows rapidly with the number of antennas and users, making real-time implementation challenging for large-scale HAPS systems.

Recent studies have explored HAPS-enabled and NTN communications, focusing primarily on coverage analysis, channel modeling, beamforming design, and network architecture \cite{Kanani2025,Xi2016,Lian2019,Zong2019,Tashiro2023}. However, robust energy-efficient beamforming for massive MIMO HAPS systems remains relatively unexplored. Moreover, conventional terrestrial formulations are not directly applicable to HAPS systems because HAPS channels exhibit geometry-dependent propagation, LoS-dominant Rician fading, and elevation-dependent interference characteristics, as captured by 3GPP NTN channel models.

A central challenge in HAPS massive MIMO systems is the acquisition of reliable CSI. Owing to the large coverage footprint, feedback delay, user mobility, finite-rate quantization, and packet loss, the transmitter operates with imperfect and temporally outdated CSI. These impairments become particularly severe at high carrier frequencies, where Doppler-induced channel aging and feedback errors can substantially degrade beamforming accuracy \cite{Dahrouj2023}. Unlike terrestrial systems, where CSI errors are often modeled using isotropic or norm-bounded uncertainty regions, HAPS channels exhibit structured uncertainty governed by LoS geometry, elevation-dependent propagation, and low-dimensional angular dynamics \cite{Zhu2022}. Thus, robust beamforming strategies for HAPS systems must account for the statistical and geometric structure of CSI degradation.

Existing robust optimization approaches for imperfect-CSI massive MIMO systems typically rely on semidefinite relaxation (SDR), successive convex approximation, or worst-case convex reformulations \cite{Jafri2022}. Although these methods provide useful theoretical guarantees, their complexity scales poorly with the number of antennas and users, making real-time implementation difficult for large-scale HAPS deployments. Learning-based methods can simplify complex optimization into low-complexity online policies \cite{Qing2023}, but most existing approaches treat CSI as an unstructured high-dimensional input and ignore the physics of LoS-dominant HAPS channels. Thus, data-driven black-box models may produce poorly calibrated uncertainty estimates and unreliable beamforming decisions under imperfect CSI. These limitations motivate \textit{physics-informed learning} frameworks \cite{Karniadakis2021} that incorporate geometric channel structure and uncertainty statistics directly into the learning and control process.

\subsection{Contributions}

Motivated by these limitations, this paper develops a physics-informed uncertainty-aware learning framework for robust energy-efficient beamforming in HAPS massive MIMO systems serving ground UEs under imperfect CSI. Specifically, we first construct a geometry-aware CSI impairment model that captures Doppler-induced channel aging, finite-rate BFI quantization, packet loss, and estimation noise. These impairments are mapped into structured tangent-space CDI uncertainty sets that respect the unit-norm geometry of channel direction information. We then design a physics-informed variational autoencoder (VAE) \cite{Pinheiro2021} to refine imperfect CDI and produce calibrated covariance estimates on the CDI manifold. Finally, the learned uncertainty representation is embedded into a multi-agent deep deterministic policy gradient (MADDPG)-based beamforming framework with centralized training \cite{lowe2017multi}, decentralized execution, and a differentiable power-projection layer for feasible real-time robust precoder generation.

The key contributions of this paper are outlined as follows:
\begin{itemize}
    \item We develop a geometry-aware HAPS massive MIMO model that captures elevation-dependent Rician fading, mobility-induced channel aging, finite-rate feedback quantization, packet loss, and estimation noise. These impairments are mapped into tangent-space CDI uncertainty sets consistent with the unit-norm structure of channel direction information.

    \item We design a physics-informed, manifold-aware VAE conditioned on the LoS steering vector and Rician factor to enhance imperfect CDI and produce calibrated tangent-space uncertainty estimates.

    \item We formulate a robust energy-efficiency maximization problem with worst-case SINR constraints, linking HAPS-specific CSI degradation to approximate probabilistic QoS guarantees through learned ellipsoidal uncertainty sets.

    \item We develop an uncertainty-aware MADDPG beamforming framework with centralized training and decentralized execution. A differentiable projection layer enforces the aggregate transmit-power constraint and enables feasible real-time precoder generation.

    \item We validate the proposed framework under practical HAPS impairments, including feedback delay, mobility, Doppler effects, packet loss, imperfect CSI, and user-density scaling. Numerical results demonstrate improved energy efficiency, SINR robustness, outage control, convergence behavior, and online computational efficiency over imperfect-CSI, SDR-based, and no-VAE baselines.
\end{itemize}

The proposed framework bridges model-based robust optimization and physics-informed generative learning, providing a computationally tractable solution for uncertainty-aware energy-efficient beamforming in HAPS massive MIMO systems.

\begin{figure}[t]
    \centering
    \includegraphics[width=0.5\textwidth]{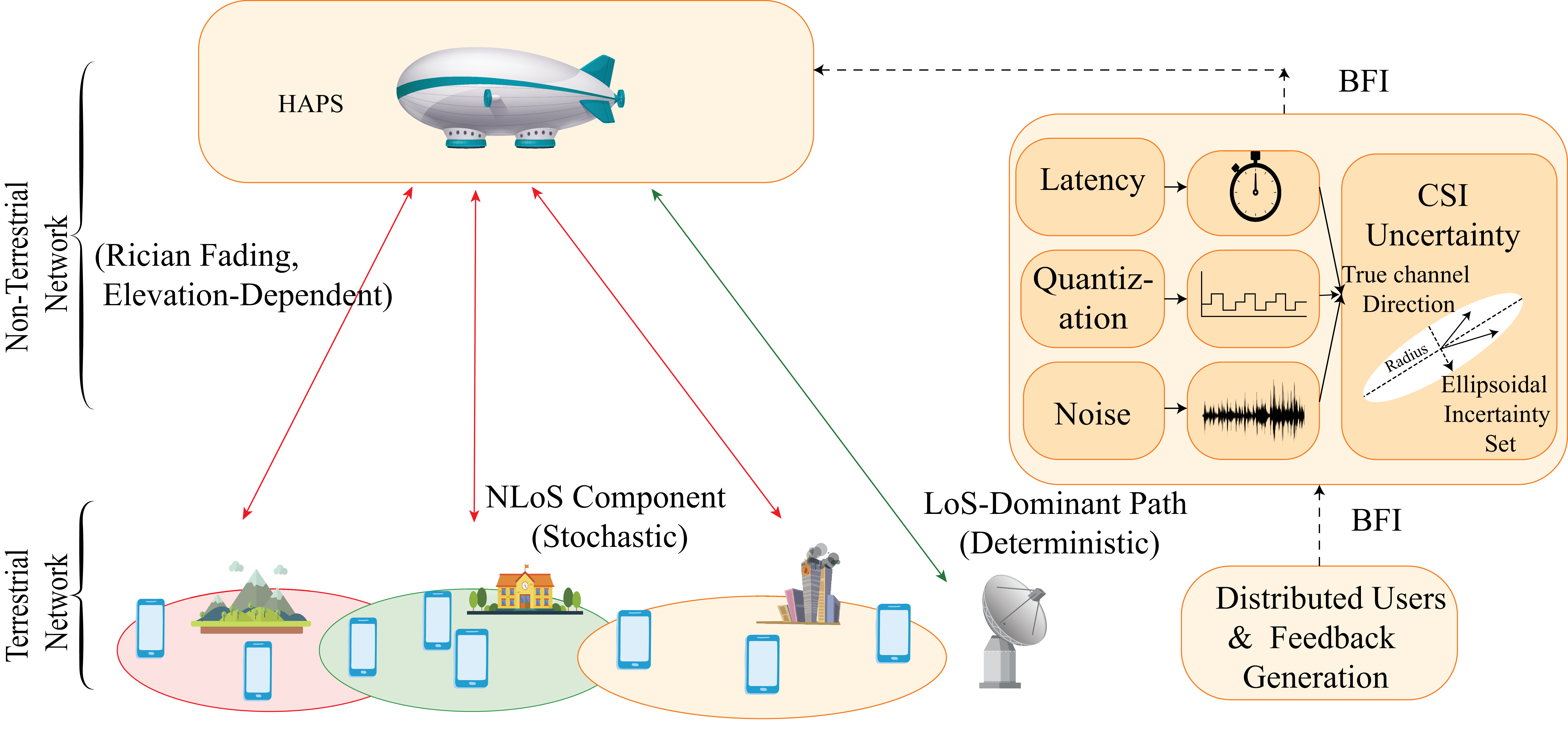}
    \caption{HAPS-assisted massive MIMO downlink. A stratospheric HAPS
    equipped with a UPA serves multiple single-antenna ground UEs over
    an elevation-dependent Rician channel; finite-rate Grassmannian
    feedback over a long round-trip link produces structured CSI
    degradation that is captured by a tangent-space ellipsoidal
    uncertainty set on the unit-norm CDI manifold.}
    \label{haps}
\end{figure}

\section{System Model}
\label{sec:system_model}
We consider the downlink of a HAPS-assisted massive MIMO network in which a quasi-stationary HAPS at altitude \(H\) serves \(\mathcal{K}=\{1,\ldots,K\}\) single-antenna ground UEs. The HAPS is equipped with an \(M\)-element uniform planar array (UPA), where \(M=M_xM_y\). The HAPS ground projection is located at \((x_H,y_H)\), while UE~\(k\) occupies the time-varying horizontal position \((x_k(t),y_k(t))\), as illustrated in Fig.~\ref{haps}. The horizontal HAPS--UE distance, slant range, elevation angle, and azimuth angle are given by
\begin{equation}
r_k(t)=\sqrt{(x_k(t)-x_H)^2+(y_k(t)-y_H)^2},
\label{eq:horizontal_distance}
\end{equation}
\begin{equation}
d_k(t)=\sqrt{r_k^2(t)+H^2},
\label{eq:slant_distance}
\end{equation}
\begin{equation}
\theta_k(t)=\arcsin\!\left(\frac{H}{d_k(t)}\right),
\label{eq:elevation_angle}
\end{equation}
\begin{equation}
\varphi_k(t)=\operatorname{atan2}\!\left(y_k(t)-y_H,\,x_k(t)-x_H\right).
\label{eq:azimuth_angle}
\end{equation}
The pair \((\theta_k(t),\varphi_k(t))\) determines the direction of departure from the HAPS array.
The UPA is assumed to be mounted on the nadir-facing side of the HAPS and to lie in the horizontal \(x\)--\(y\) plane. Under half-wavelength spacing, the normalized UPA steering vector is
\begin{equation}
\mathbf{a}(\theta_k,\varphi_k)
=
\mathbf{a}_x(\theta_k,\varphi_k)
\otimes
\mathbf{a}_y(\theta_k,\varphi_k),
\label{eq:upa_steering}
\end{equation}
where
\begin{equation}
[\mathbf{a}_x]_m
=
\frac{1}{\sqrt{M_x}}
e^{j\pi(m-1)\cos\theta_k\cos\varphi_k},
\quad m=1,\ldots,M_x,
\end{equation}
and
\begin{equation}
[\mathbf{a}_y]_n
=
\frac{1}{\sqrt{M_y}}
e^{j\pi(n-1)\cos\theta_k\sin\varphi_k},
\quad n=1,\ldots,M_y.
\end{equation}
Hence, \(\|\mathbf{a}(\theta_k,\varphi_k)\|_2^2=1\).
The HAPS--UE large-scale gain combines free-space loss with an elevation-dependent atmospheric path-length factor~\cite{ITU676,3gpp38811}:
\begin{equation}
\beta_k(t)
=
\left(\frac{c}{4\pi f_c d_k(t)}\right)^2
10^{-A(f_c)h_{\mathrm{atm}}/(10\sin\theta_k(t))},
\label{eq:pathloss}
\end{equation}
where \(c\) is the speed of light, \(f_c\) is the carrier frequency, \(A(f_c)\) is the gaseous attenuation coefficient in dB/km, and \(h_{\mathrm{atm}}\) is the equivalent atmospheric scale height. The factor \(1/\sin\theta_k(t)\) accounts for the elevation-dependent atmospheric path length under the planar-stratified approximation.

The small-scale fading follows an elevation-dependent Rician model commonly adopted for NTNs and HAPS channels~\cite{3gpp38811}:
\begin{equation}
\mathbf{h}_k(t)
=
\sqrt{\beta_k(t)}
\left(
\sqrt{\frac{\kappa_k(t)}{\kappa_k(t)+1}}\,
\bar{\mathbf{h}}_k(t)
+
\sqrt{\frac{1}{\kappa_k(t)+1}}\,
\tilde{\mathbf{h}}_k(t)
\right),
\label{eq:rician_channel}
\end{equation}
where \(\kappa_k(t)\) is the linear-scale Rician factor and
\(\tilde{\mathbf{h}}_k(t)\sim\mathcal{CN}(\mathbf{0},\mathbf{I}_M)\). The deterministic LoS component is
\begin{equation}
\bar{\mathbf{h}}_k(t)
=
\sqrt{M}\,
e^{-j2\pi f_c d_k(t)/c}
\mathbf{a}(\theta_k(t),\varphi_k(t)).
\label{eq:LoS}
\end{equation}
The factor \(\sqrt{M}\) ensures \(\|\bar{\mathbf{h}}_k(t)\|_2^2=M\), yielding
\(\mathbb{E}\{\|\mathbf{h}_k(t)\|_2^2\}=M\beta_k(t)\).

The Rician factor is modeled as an elevation-dependent parameter \cite{Lin2026}:
\begin{equation}
\kappa_k[\mathrm{dB}]
=
\kappa_0+\kappa_1\theta_k+\kappa_2\theta_k^2,
\label{eq:rician_k_factor}
\end{equation}
where \(\kappa_0\), \(\kappa_1\), and \(\kappa_2\) are environment-dependent coefficients, and \(\theta_k\) is expressed in degrees in this empirical model. The linear-scale factor used in \eqref{eq:rician_channel} is obtained as
\(\kappa_k=10^{\kappa_k[\mathrm{dB}]/10}\).

\subsection{Mobility and Temporal Channel Evolution}

For UE mobility with speed \(v_k(t)\) and heading \(\psi_{v,k}(t)\) in the horizontal plane, the Doppler shift is given by the projection of the UE velocity onto the horizontal component of the HAPS--UE propagation direction:
\begin{equation}
f_{D,k}(t)
=
\frac{v_k(t)}{\lambda}
\cos\theta_k(t)
\cos\!\big(\psi_{v,k}(t)-\varphi_k(t)\big),
\label{eq:doppler}
\end{equation}
where \(\lambda=c/f_c\). The maximum Doppler frequency is \(f_{D,\max}=v_{\max}f_c/c\). Following the Clarke--Jakes temporal-correlation model~\cite{Clarke1968,Jakes1994}, the coherence time is approximated as \(T_c\approx0.423/f_{D,\max}\). Although this expression is originally derived for isotropic scattering, it provides a conservative reference time scale for the scattered component of the Rician HAPS channel.

Over short prediction horizons and assuming \(f_{D,k}\) remains approximately constant, the LoS component evolves mainly through a phase rotation, since the angular displacement over one coherence interval is negligible relative to the UPA beamwidth under the considered HAPS geometry:
\begin{equation}
\bar{\mathbf{h}}_k(t)
\approx
e^{j2\pi f_{D,k}(t-t_0)}
\bar{\mathbf{h}}_k(t_0).
\label{eq:LoS_evo}
\end{equation}
The scattered component follows a first-order Gauss--Markov recursion whose correlation coefficient is matched to the Clarke--Jakes temporal autocorrelation model~\cite{Clarke1968,Jakes1994}:
\begin{equation}
\tilde{\mathbf{h}}_k[n]
=
\alpha_k\tilde{\mathbf{h}}_k[n-1]
+
\sqrt{1-\alpha_k^2}\,\mathbf{e}_k[n],
\quad
\alpha_k=J_0(2\pi f_{D,k}T_s),
\label{eq:gauss_markov_channel}
\end{equation}
where \(T_s\) is the channel sampling interval, \(J_0(\cdot)\) is the zeroth-order Bessel function arising from the Clarke--Jakes temporal autocorrelation model, and \(\mathbf{e}_k[n]\sim\mathcal{CN}(\mathbf{0},\mathbf{I}_M)\). The \(\tau\)-step prediction-error covariance of the scattered component is \((1-\alpha_k^{2\tau})\mathbf{I}_M\).

\subsection{Beamforming Feedback Information Impairments}

Let
\begin{equation}
\mathbf{g}_k[n]\triangleq \frac{\mathbf{h}_k[n]}{\|\mathbf{h}_k[n]\|_2}
\end{equation}
denote the CDI of UE~\(k\) at slot~\(n\), represented up to an irrelevant common phase. Owing to feedback delay, the transmitter observes the delayed CDI
\begin{equation}
\mathbf{g}_k^{(\mathrm{del})}[n]=\mathbf{g}_k[n-\tau],
\quad
\tau=\left\lceil\frac{T_{\mathrm{delay}}}{T_s}\right\rceil .
\end{equation}
Finite-rate Grassmannian quantization is modeled as
\begin{equation}
\hat{\mathbf{g}}_k^{(q)}[n]
=
\sqrt{1-\sigma_{q,k}^2}\,
\mathbf{g}_k^{(\mathrm{del})}[n]
+
\sigma_{q,k}\mathbf{n}_{q,k}[n],
\label{eq:quantized_cdi}
\end{equation}
where \(\mathbf{n}_{q,k}[n]\) is a unit-norm error vector orthogonal to \(\mathbf{g}_k^{(\mathrm{del})}[n]\). Allocating the feedback budget \(B_{\mathrm{LoS}}\) to the dominant LoS subspace of effective dimension \(d_{\mathrm{eff}}\), the random vector quantization bound gives
\begin{equation}
\sigma_{q,k}^2
=
\min\!\left\{
1,\,
\frac{1}{\kappa_k+1}
+
\frac{\kappa_k}{\kappa_k+1}
2^{-B_{\mathrm{LoS}}/(d_{\mathrm{eff}}-1)}
\right\}.
\label{eq:sigma_q}
\end{equation}

Packet loss on the feedback link is captured as
\begin{equation}
\hat{\mathbf{g}}_k[n]
=
\xi_k[n]\hat{\mathbf{g}}_k^{(q)}[n]
+
\big(1-\xi_k[n]\big)\hat{\mathbf{g}}_k[n-1],
\label{eq:packet_loss_cdi}
\end{equation}
where \(\xi_k[n]\in\{0,1\}\) is Bernoulli with retention probability \(1-p_{\mathrm{loss}}\). Lost packets cause the transmitter to retain the most recently received CDI.

\subsection{Downlink Signal Model under Imperfect CSI}

The HAPS transmits
\begin{equation}
\mathbf{x}[n]
=
\sum_{k=1}^{K}\mathbf{v}_k[n]s_k[n],
\label{eq:tx_signal}
\end{equation}
where \(s_k[n]\) is the unit-power data symbol for UE~\(k\), and \(\mathbf{v}_k[n]\in\mathbb{C}^{M}\) is the full downlink precoder including both beam direction and transmit power. The aggregate power constraint is
\begin{equation}
\sum_{k=1}^{K}\|\mathbf{v}_k[n]\|_2^2 \le P_{\max}.
\label{eq:power_constraint}
\end{equation}

For notational simplicity, slot indices are suppressed in the following expressions. The received signal at UE~\(k\) is
\begin{equation}
y_k
=
\mathbf{h}_k^H\mathbf{v}_k s_k
+
\sum_{j\neq k}\mathbf{h}_k^H\mathbf{v}_j s_j
+
n_k,
\label{eq:received_signal}
\end{equation}
where \(n_k\sim\mathcal{CN}(0,\sigma_n^2)\). Under perfect CSI, the instantaneous SINR is
\begin{equation}
\gamma_k
=
\frac{|\mathbf{h}_k^H\mathbf{v}_k|^2}
{\sum_{j\neq k}|\mathbf{h}_k^H\mathbf{v}_j|^2+\sigma_n^2}.
\label{eq:sinr_perfect_csi}
\end{equation}

Under imperfect CSI, we write
\begin{equation}
\mathbf{h}_k=\hat{\mathbf{h}}_k+\Delta\mathbf{h}_k,
\label{eq:channel_error_model}
\end{equation}
with channel-error covariance
\(\mathbf{C}_{h,k}=\mathbb{E}\{\Delta\mathbf{h}_k\Delta\mathbf{h}_k^H\}\). The corresponding approximate average SINR is
\begin{equation}
\bar{\gamma}_k
\approx
\frac{|\hat{\mathbf{h}}_k^H\mathbf{v}_k|^2}
{\sum_{j\neq k}|\hat{\mathbf{h}}_k^H\mathbf{v}_j|^2
+\sum_{j=1}^{K}\mathbf{v}_j^H\mathbf{C}_{h,k}\mathbf{v}_j
+\sigma_n^2}.
\label{eq:sinr_imperfect_csi}
\end{equation}
The structured form of \(\mathbf{C}_{h,k}\) on the unit-norm CDI manifold is derived next.

\subsection{Structured CSI Uncertainty Set}
\label{subsec:csi_error}
Let \(\boldsymbol{\delta}_k=\mathbf{g}_k-\hat{\mathbf{g}}_k\) denote the CDI error. Since both \(\mathbf{g}_k\) and \(\hat{\mathbf{g}}_k\) are unit-norm CDI vectors, the error can be locally represented in the tangent space at \(\hat{\mathbf{g}}_k\), consistent with Grassmannian limited-feedback channel-direction modeling~\cite{Love2003,Absil2008}. We model this error as zero-mean Gaussian with covariance
\begin{equation}
\mathbf{C}_{g,k}
=
\mathbf{C}_{g,k}^{(\mathrm{age})}
+
\sigma_{q,k}^2\boldsymbol{\Pi}_k^\perp
+
\sigma_{e,k}^2\mathbf{I}_M,
\label{eq:error_covariance}
\end{equation}
where \(\sigma_{e,k}^2\) models residual estimation noise and
\(\boldsymbol{\Pi}_k^\perp=\mathbf{I}_M-\hat{\mathbf{g}}_k\hat{\mathbf{g}}_k^H\) is the tangent-space projection.

The aging covariance is decomposed into LoS angular perturbation and NLoS decorrelation terms:
\begin{equation}
\mathbf{C}_{g,k}^{(\mathrm{age})}
=
\frac{\kappa_k}{\kappa_k+1}
\mathbf{J}_{a,k}\boldsymbol{\Sigma}_{\theta\varphi,k}\mathbf{J}_{a,k}^H
+
\frac{1-\alpha_k^{2\tau}}{\kappa_k+1}
\boldsymbol{\Pi}_k^\perp,
\label{eq:aging_covariance}
\end{equation}
where
\(\mathbf{J}_{a,k}=[\partial\mathbf{a}/\partial\theta_k,\partial\mathbf{a}/\partial\varphi_k]\), and
\(\boldsymbol{\Sigma}_{\theta\varphi,k}\) is the delayed angular prediction-error covariance. The common LoS propagation phase does not affect beam-domain power gains and is therefore excluded from the CDI uncertainty model.
Writing \(\mathbf{h}_k=\sqrt{\eta_k}\mathbf{g}_k\), where \(\eta_k=\|\mathbf{h}_k\|_2^2\), the channel-domain covariance is approximated as
\begin{equation}
\mathbf{C}_{h,k}\approx \hat{\eta}_k\mathbf{C}_{g,k}.
\label{eq:cdi_to_channel_covariance}
\end{equation}
This approximation relies on a separation of timescales: the channel gain is tracked through slow-timescale CQI feedback, whereas CDI uncertainty is dominated by faster Doppler decorrelation, finite-rate quantization, feedback delay, and packet loss.

The CDI uncertainty set is defined as the tangent-space ellipsoid~\cite{BenTal2009,Bertsimas2011}
\begin{equation}
\mathcal{U}_k
=
\left\{
\boldsymbol{\delta}\in\mathbb{C}^{M}:
\boldsymbol{\delta}^H
\left(\mathbf{C}_{g,k}+\epsilon\mathbf{I}_M\right)^{-1}
\boldsymbol{\delta}
\le
\varepsilon_k^2
\right\},
\label{eq:uncertainty_set}
\end{equation}
where \(\epsilon>0\) is a small regularization parameter. The radius is selected from the chi-squared quantile
\begin{equation}
\varepsilon_k^2
\approx
\chi_{2\nu_k}^{-1}(1-\epsilon_{\mathrm{out}}),
\label{eq:uncertainty_radius}
\end{equation}
where
\begin{equation}
\nu_k
=
\left\lfloor
\frac{\left[\mathrm{tr}(\mathbf{C}_{g,k})\right]^2}
{\mathrm{tr}(\mathbf{C}_{g,k}^2)}
\right\rfloor
\end{equation}
is the eigenvalue participation ratio of \(\mathbf{C}_{g,k}\) \cite{Nadler_2008}. The set satisfies the approximate coverage relation
\begin{equation}
\Pr\!\left(\mathbf{g}_k\in\hat{\mathbf{g}}_k+\mathcal{U}_k\right)
\approx
1-\epsilon_{\mathrm{out}},
\label{eq:coverage_probability}
\end{equation}
which connects the statistical CSI impairment model to a deterministic uncertainty set suitable for robust beamforming.

\section{Robust Energy-Efficient Beamforming Formulation}
\label{sec:robust_formulation}

This section formulates the robust downlink beamforming problem using the structured CDI uncertainty set \(\mathcal{U}_k\) in~\eqref{eq:uncertainty_set}. The design is driven by two coupled requirements: (i) each UE must satisfy a reliability-aware QoS constraint under HAPS-specific CSI degradation, and (ii) the HAPS transmit power must be used efficiently under strict payload energy constraints.

\subsection{Probabilistic QoS and Worst-Case SINR Constraint}

Each UE requires a minimum spectral efficiency \(R_k^{(\mathrm{th})}\), equivalently an SINR threshold
\begin{equation}
\gamma_k^{(\mathrm{th})}=2^{R_k^{(\mathrm{th})}}-1 .
\end{equation}
Because instantaneous QoS guarantees are difficult to maintain under feedback delay, channel aging, and finite-rate BFI, service reliability is enforced probabilistically as
\begin{equation}
\Pr\!\left(
\gamma_k < \gamma_k^{(\mathrm{th})}
\,\big|\,
\hat{\mathbf{g}}_k,\{\mathbf{v}_j\}_{j=1}^{K}
\right)
\le
\epsilon_{\mathrm{out}},
\quad \forall k\in\mathcal{K}.
\label{eq:prob_qos}
\end{equation}

Using the CDI uncertainty set \(\mathcal{U}_k\), the chance constraint in~\eqref{eq:prob_qos} is conservatively replaced by the deterministic worst-case SINR requirement
\begin{equation}
\min_{\boldsymbol{\delta}_k\in\mathcal{U}_k}
\gamma_k\!\left(
\{\mathbf{v}_j\}_{j=1}^{K},
\hat{\mathbf{g}}_k+\boldsymbol{\delta}_k
\right)
\ge
\gamma_k^{(\mathrm{th})},
\quad \forall k\in\mathcal{K}.
\label{eq:robust_constraint}
\end{equation}
This constraint requires the SINR target to hold for all CDI perturbations inside the calibrated tangent-space ellipsoid. If \(\mathcal{U}_k\) achieves coverage probability \(1-\epsilon_{\mathrm{out}}\), then satisfying \eqref{eq:robust_constraint} implies an approximate probabilistic QoS guarantee with outage probability bounded by \(\epsilon_{\mathrm{out}}\). This interpretation relies on the tangent-space Gaussian approximation and the chi-squared radius calibration in~\eqref{eq:uncertainty_radius}; its empirical accuracy is evaluated in the simulation section.

Classical solutions to~\eqref{eq:prob_qos} often rely on Bernstein-type inequalities, S-procedure reformulations, or  SDR, which yield tractable but conservative formulations with per-slot complexity that scales poorly with the number of antennas and users~\cite{9857564}. In contrast, the proposed framework embeds the learned uncertainty representation into the online beamforming policy, avoiding the need to solve a high-dimensional robust optimization problem at every coherence interval.

\subsection{Energy-Efficiency Objective and Robust Optimization Problem}

The system energy efficiency is defined as the ratio between the achievable sum rate and the total consumed power:
\begin{equation}
\eta_{\mathrm{EE}}
=
\frac{
W\sum_{k=1}^{K}\log_2\!\left(1+\bar{\gamma}_k\right)
}
{
\eta_{\mathrm{PA}}^{-1}\sum_{k=1}^{K}\|\mathbf{v}_k\|_2^2
+
P_{\mathrm{fix}}
}
\quad [\mathrm{bit/J}],
\label{eq:energy_efficiency}
\end{equation}
where \(W\) is the system bandwidth, \(\eta_{\mathrm{PA}}\in(0,1]\) is the power-amplifier efficiency, \(P_{\mathrm{fix}}\) denotes the fixed circuit and payload-related power consumption, and \(\mathbf{v}_k\) is the full downlink precoder of UE~\(k\), including both beam direction and transmit power.

The robust energy-efficient beamforming problem is formulated as
\begin{subequations}
\label{eq:P1}
\begin{align}
\mathcal{P}_1:
\max_{\{\mathbf{v}_k\}_{k=1}^{K}}
\quad
&
\eta_{\mathrm{EE}}
\label{eq:P1_obj}
\\
\mathrm{s.t.}\quad
&
\min_{\boldsymbol{\delta}_k\in\mathcal{U}_k}
\gamma_k\!\left(
\{\mathbf{v}_j\}_{j=1}^{K},
\hat{\mathbf{g}}_k+\boldsymbol{\delta}_k
\right)
\ge
\gamma_k^{(\mathrm{th})},
\; \forall k\in \!\mathcal{K},
\label{eq:P1_robust_qos}
\\
&
\sum_{k=1}^{K}\|\mathbf{v}_k\|_2^2
\le
P_{\max}.
\label{eq:P1_power}
\end{align}
\end{subequations}

Problem~\(\mathcal{P}_1\) is challenging for three reasons. First, the objective in~\eqref{eq:P1_obj} is fractional and nonconvex due to the coupled SINR terms. Second, the robust QoS constraints in~\eqref{eq:P1_robust_qos} are semi-infinite because they must hold for all CDI perturbations inside the uncertainty sets. Third, multiuser interference couples all precoders, preventing per-user decomposition. Conventional Dinkelbach--SDR or SCA-based approaches~\cite{Dinkelbach1967,Luo2010,Scutari2017} can provide useful benchmarks but require repeated high-dimensional optimization, which is computationally prohibitive for real-time HAPS beamforming. The next section develops a physics-informed VAE--MADDPG framework that maps this robust beamforming problem into a low-complexity online policy while preserving the uncertainty-aware structure of~\(\mathcal{P}_1\).

\section{Physics-Informed VAE--MADDPG Beamforming Framework}
\label{sec:framework}
\begin{figure*}[t]
    \centering
    \includegraphics[width=0.85\textwidth]{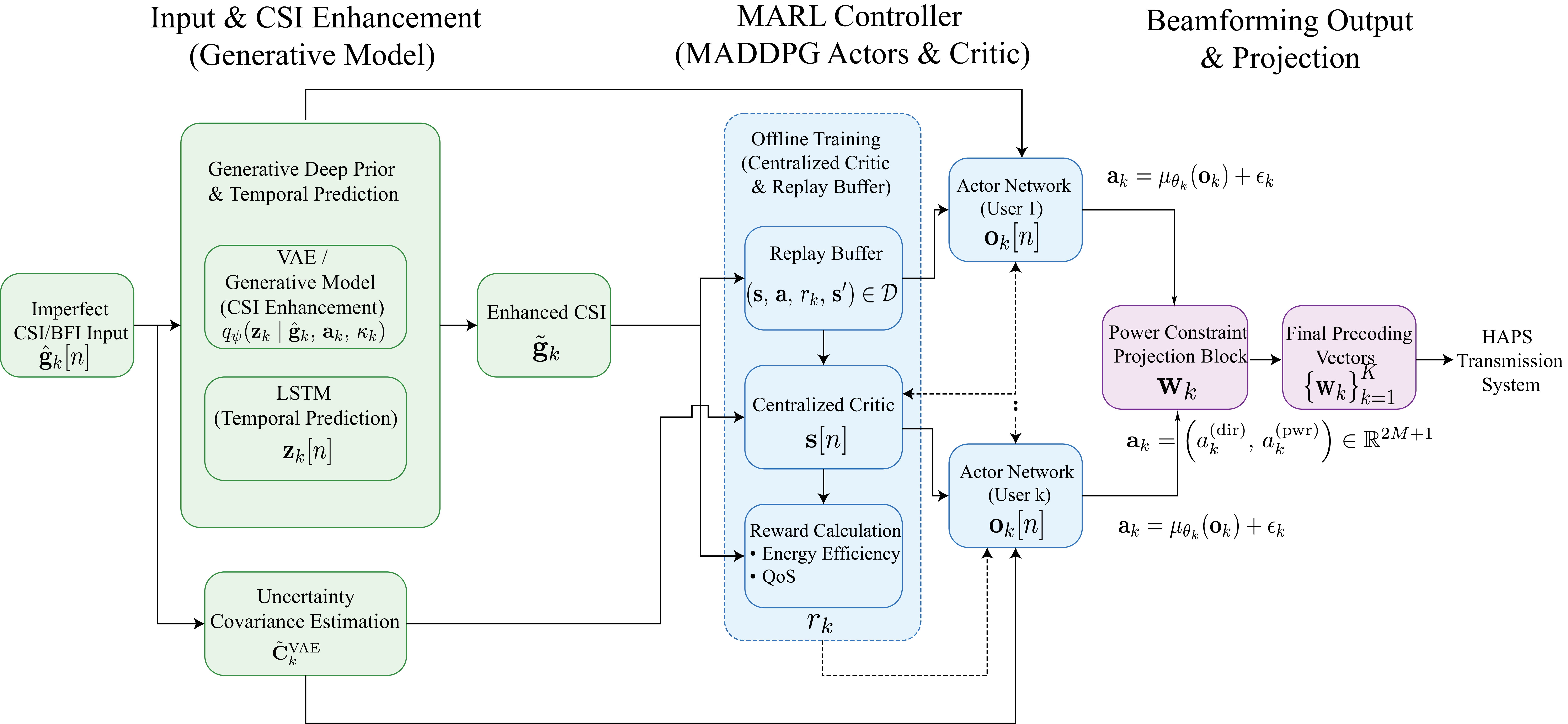}
    \caption{Proposed physics-informed uncertainty-aware beamforming framework for HAPS massive MIMO systems}
    \label{fig:architecture}
\end{figure*}

This section develops the proposed physics-informed uncertainty-aware learning framework for solving \(\mathcal{P}_1\). As shown in Fig.~\ref{fig:architecture}, the proposed framework combines a physics-informed CDI enhancement stage with an uncertainty-aware MADDPG beamforming controller. First, a VAE refines the imperfect CDI obtained from delayed, quantized, and possibly lost BFI, while producing a calibrated tangent-space covariance. Second, this covariance is used as an uncertainty descriptor of CSI reliability. Third, the enhanced CDI and uncertainty descriptor are embedded into a MADDPG policy, whose output is projected onto the feasible transmit-power set to generate robust downlink precoders.

Throughout this section, the action of agent~\(k\) is denoted by
\(\boldsymbol{\alpha}_k\), to avoid notation collision with the UPA steering vector \(\mathbf{a}(\theta_k,\varphi_k)\). The tangent-space projection matrix associated with the enhanced CDI estimate is
\begin{equation}
\boldsymbol{\Pi}_k^\perp
=
\mathbf{I}_M-\tilde{\mathbf{g}}_k\tilde{\mathbf{g}}_k^H .
\end{equation}

\subsection{Physics-Informed Manifold-Aware Uncertainty Learning}

For LoS-dominant Rician channels, CDI realizations concentrate around the steering direction \(\mathbf{a}(\theta_k,\varphi_k)\) on a low-dimensional manifold determined by the angular channel geometry. The proposed VAE exploits this structure to learn enhanced CDI representations and calibrated tangent-space uncertainty estimates under HAPS-specific feedback impairments.

\subsubsection{Encoder-decoder with unit-sphere projection}

The encoder produces the latent Gaussian posterior
\begin{equation}
q_{\boldsymbol{\psi}}
(
\mathbf{z}_k
\mid
\hat{\mathbf{g}}_k,
\mathbf{a}_k,
\kappa_k
)
=
\mathcal{N}
(
\boldsymbol{\mu}_z,
\mathrm{diag}(\boldsymbol{\sigma}_z^2)
),
\label{eq:encoder_dist}
\end{equation}
where
\begin{equation}
[
\boldsymbol{\mu}_z,
\log \boldsymbol{\sigma}_z^2
]
=
f_{\boldsymbol{\psi}}^{(\mathrm{enc})}
(
\hat{\mathbf{g}}_k,
\mathbf{a}_k,
\kappa_k
).
\label{eq:encoder}
\end{equation}

The decoder reconstructs the CDI distribution as
\begin{equation}
p_{\boldsymbol{\omega}}
(
\mathbf{g}_k
\mid
\mathbf{z}_k,
\mathbf{a}_k,
\kappa_k
)
=
\mathcal{CN}
(
\boldsymbol{\mu}_g,
\mathrm{diag}(\boldsymbol{\sigma}_g^2)
),
\label{eq:decoder_dist}
\end{equation}
with
\begin{equation}
[
\boldsymbol{\mu}_g,
\boldsymbol{\sigma}_g^2
]
=
g_{\boldsymbol{\omega}}
(
\mathbf{z}_k,
\mathbf{a}_k,
\kappa_k
).
\label{eq:decoder}
\end{equation}

Conditioning the VAE on the steering vector and Rician factor embeds the dominant geometric channel structure into the latent representation, allowing the latent variables to model stochastic feedback impairments and residual channel uncertainty. Since CDI vectors lie on the unit hypersphere, the reconstructed mean is projected as
\begin{equation}
\tilde{\mathbf{g}}_k
=
\frac{\boldsymbol{\mu}_g}
{\|\boldsymbol{\mu}_g\|_2}.
\label{eq:sphere_projection}
\end{equation}

The corresponding tangent-space covariance is obtained through first-order Jacobian propagation:
\begin{equation}
\mathbf{C}_k^{(\mathrm{proj})}
=
\frac{
\boldsymbol{\Pi}_k^\perp
\,
\mathrm{diag}(\boldsymbol{\sigma}_g^2)
\,
\boldsymbol{\Pi}_k^\perp
}
{
\|\boldsymbol{\mu}_g\|_2^2
}.
\label{eq:proj_covariance}
\end{equation}

\subsubsection{Physics-informed prior and training objective}

The latent prior is selected to reflect the LoS-dominant channel geometry:
\begin{equation}
p(\mathbf{z}_k|\kappa_k)
=
\mathcal{N}\!\left(
\mathbf{0},
c_p(\kappa_k+1)^{-1}\mathbf{I}_{d_z}
\right),
\label{eq:prior}
\end{equation}
where \(d_z\) is the latent dimension and \(c_p\) is a scaling constant. This prior contracts as the Rician factor increases, reflecting reduced channel-direction uncertainty under stronger LoS dominance.

The VAE is trained by minimizing the negative evidence lower bound (ELBO), following the standard variational inference principle~\cite{kingma2013auto,chen2018vari}:
\begin{equation}
\mathcal{L}(\boldsymbol{\psi},\boldsymbol{\omega})
=
-\mathcal{L}_{\mathrm{ELBO}}
+
\lambda_{\mathrm{LoS}}
\mathcal{L}_{\mathrm{LoS}},
\label{eq:total_loss}
\end{equation}
where \begin{equation}
\begin{aligned}
\mathcal{L}_{\mathrm{ELBO}}
&=
\mathbb{E}_{q_{\boldsymbol{\psi}}}
\!\left[
\log
p_{\boldsymbol{\omega}}(\mathbf{g}_k|\mathbf{z}_k)
\right]
\\
&\quad
-
D_{\mathrm{KL}}
\!\left(
q_{\boldsymbol{\psi}}
(\mathbf{z}_k|\hat{\mathbf{g}}_k,\mathbf{a}_k,\kappa_k)
\;\|\;
p(\mathbf{z}_k|\kappa_k)
\right).
\end{aligned}
\end{equation}
The regularizer discourages reconstructed CDI components that are inconsistent with the dominant LoS steering subspace.

\subsubsection{Learned ellipsoidal uncertainty set}
The final learned covariance combines the projected decoder uncertainty and the latent uncertainty propagated through the decoder:
\begin{equation}
\tilde{\mathbf{C}}_k^{(\mathrm{VAE})}
=
\frac{
\boldsymbol{\Pi}_k^\perp
\left(
\mathrm{diag}(\boldsymbol{\sigma}_g^2)
+
\mathbf{J}_g
\mathrm{diag}(\boldsymbol{\sigma}_z^2)
\mathbf{J}_g^H
\right)
\boldsymbol{\Pi}_k^\perp
}
{
\|\boldsymbol{\mu}_g\|_2^2
},
\label{eq:vae_covariance}
\end{equation}
where
\[
\mathbf{J}_g
=
\left.
\frac{\partial \boldsymbol{\mu}_g}{\partial \mathbf{z}}
\right|_{\mathbf{z}=\boldsymbol{\mu}_z}
\]
is the decoder Jacobian evaluated at the posterior mean. The learned covariance parameterizes the ellipsoidal set
\begin{equation}
\tilde{\mathcal{U}}_k
=
\left\{
\boldsymbol{\delta}\in\mathbb{C}^{M}:
\boldsymbol{\delta}^H
\left(\tilde{\mathbf{C}}_k^{(\mathrm{VAE})}+\epsilon\mathbf{I}_M\right)^{-1}
\boldsymbol{\delta}
\le
\chi_{2\nu_k}^{-1}(1-\epsilon_{\mathrm{out}})
\right\},
\label{eq:learned_set}
\end{equation}
where \(\epsilon>0\) is a small regularization parameter and \(\nu_k\) is the effective dimension defined in~\eqref{eq:uncertainty_radius}. The learned set \(\tilde{\mathcal{U}}_k\) refines the model-based set \(\mathcal{U}_k\) by replacing the analytical covariance with the VAE-calibrated tangent-space covariance.
\begin{algorithm}[t]
\caption{Physics-informed VAE training and inference for CDI uncertainty estimation}
\label{alg:vae_csi}
\begin{algorithmic}[1]
\REQUIRE Training CDI pairs \(\{(\hat{\mathbf{g}}_k,\mathbf{g}_k,\mathbf{a}_k,\kappa_k)\}\), encoder \(f_{\boldsymbol{\psi}}^{(\mathrm{enc})}\), decoder \(g_{\boldsymbol{\omega}}\)
\ENSURE Enhanced CDI \(\tilde{\mathbf{g}}_k\) and covariance \(\tilde{\mathbf{C}}_k^{(\mathrm{VAE})}\)

\STATE \textbf{Offline training phase}
\FOR{each training epoch}
    \STATE Sample a mini-batch of imperfect and reference CDI pairs
    \FOR{each UE sample \(k\)}
        \STATE Encode: $[\boldsymbol{\mu}_z,\log\boldsymbol{\sigma}_z^2] \leftarrow f_{\boldsymbol{\psi}}^{(\mathrm{enc})} (\hat{\mathbf{g}}_k,\mathbf{a}_k,\kappa_k)$
        \STATE Sample latent vector using reparameterization: 
        $\mathbf{z}_k = \boldsymbol{\mu}_z + \boldsymbol{\sigma}_z\odot\boldsymbol{\epsilon},
        \quad
        \boldsymbol{\epsilon}\sim\mathcal{N}(\mathbf{0},\mathbf{I})$
        \STATE Decode: $
        [\boldsymbol{\mu}_g,\boldsymbol{\sigma}_g^2]
        \leftarrow
        g_{\boldsymbol{\omega}}(\mathbf{z}_k,\mathbf{a}_k,\kappa_k)$
        \STATE Project reconstructed mean:        $\tilde{\mathbf{g}}_k
        \leftarrow
        \boldsymbol{\mu}_g/\|\boldsymbol{\mu}_g\|_2$
    \ENDFOR
    \STATE Update \((\boldsymbol{\psi},\boldsymbol{\omega})\) by minimizing \(\mathcal{L}(\boldsymbol{\psi},\boldsymbol{\omega})\) in~\eqref{eq:total_loss}
\ENDFOR

\STATE \textbf{Online inference phase}
\STATE Encode \((\hat{\mathbf{g}}_k,\mathbf{a}_k,\kappa_k)\) to obtain \((\boldsymbol{\mu}_z,\boldsymbol{\sigma}_z^2)\)
\STATE Decode at posterior mean: 
$[\boldsymbol{\mu}_g,\boldsymbol{\sigma}_g^2]
\leftarrow
g_{\boldsymbol{\omega}}(\boldsymbol{\mu}_z,\mathbf{a}_k,\kappa_k)$
\STATE Project to unit sphere: 
$\tilde{\mathbf{g}}_k
\leftarrow
\boldsymbol{\mu}_g/\|\boldsymbol{\mu}_g\|_2$
\STATE Compute \(\tilde{\mathbf{C}}_k^{(\mathrm{VAE})}\) using~\eqref{eq:vae_covariance}
\RETURN \(\tilde{\mathbf{g}}_k\), \(\tilde{\mathbf{C}}_k^{(\mathrm{VAE})}\)
\end{algorithmic}
\end{algorithm}
\subsection{Uncertainty-Aware MADDPG Beamforming}

The robust beamforming problem in~\eqref{eq:P1} remains difficult to solve online due to coupled multiuser interference and semi-infinite worst-case SINR constraints. To obtain scalable real-time beamforming policies, we formulate the online beamforming task as a decentralized partially observable Markov decision process (Dec-POMDP) with centralized training and decentralized execution (CTDE)~\cite{oliehoek2016concise}.

Each agent observes only local CSI-related information, whereas the centralized critic exploits global state information during training to capture multiuser interference coupling and uncertainty interactions.

\subsubsection{State, observation, and action}

The global training state is defined as
\begin{equation}
\mathbf{s}[n]
=
\{
\tilde{\mathbf{g}}_k,
\tilde{\mathbf{C}}_k^{(\mathrm{VAE})},
\beta_k,
\gamma_k[n-1]
\}_{k=1}^{K}.
\label{eq:global_state}
\end{equation}
During online execution, agent~\(k\) observes only
\begin{equation}
\mathbf{o}_k[n]
=
\{
\tilde{\mathbf{g}}_k,
\mathrm{tr}(\tilde{\mathbf{C}}_k^{(\mathrm{VAE})})/M,
\beta_k,
\gamma_k[n-1],
\kappa_k
\}.
\label{eq:local_obs}
\end{equation}
The normalized covariance trace serves as a compact uncertainty indicator that allows the actor to adapt its beamforming aggressiveness according to CSI reliability.

The actor outputs a real-valued action vector
\begin{equation}
\boldsymbol{\alpha}_k
=
\left[
\boldsymbol{\alpha}_{k}^{(\mathrm{dir})},
\alpha_k^{(\mathrm{pwr})}
\right]
\in
\mathbb{R}^{2M+1}.
\end{equation}
The directional component is mapped into a complex beam direction:
\begin{equation}
\mathbf{d}_k
=
[\boldsymbol{\alpha}_{k}^{(\mathrm{dir})}]_{1:M}
+
j[\boldsymbol{\alpha}_{k}^{(\mathrm{dir})}]_{M+1:2M}.
\label{eq:complex_direction}
\end{equation}
The scalar power gate is defined as
\begin{equation}
q_k=\sigma(\alpha_k^{(\mathrm{pwr})}),
\quad 0\le q_k\le 1,
\end{equation}
where \(\sigma(\cdot)\) is the sigmoid function. The raw full precoder is then
\begin{equation}
\mathbf{v}_k^{(\mathrm{raw})}
=
\sqrt{P_{\max}q_k}
\frac{\mathbf{d}_k}{\|\mathbf{d}_k\|_2}.
\label{eq:raw_precoder}
\end{equation}
This construction allows the actor to control both beam direction and transmit-power usage.

\subsubsection{Differentiable power projection}

To guarantee feasibility under the aggregate transmit-power constraint, the raw precoders are jointly projected as
\begin{equation}
\mathbf{v}_k
=
\mathbf{v}_k^{(\mathrm{raw})}
\min
\left(
1,
\sqrt{
\frac{
P_{\max}
}{
\sum_{j=1}^{K}
\|\mathbf{v}_j^{(\mathrm{raw})}\|_2^2
}
}
\right).
\label{eq:power_projection}
\end{equation}
The projection preserves relative beam directions while ensuring
\begin{equation}
\sum_{k=1}^{K}\|\mathbf{v}_k\|_2^2 \le P_{\max}.
\end{equation}
Moreover, the projection operator is differentiable almost everywhere, enabling end-to-end gradient-based policy optimization.

\subsubsection{Uncertainty-aware reward}
Energy efficiency, QoS satisfaction, and power utilization are combined into the reward function as follows:
\begin{equation}
\begin{aligned}
r_k
=
&
\log\!\left(
1+
\frac{\eta_{\mathrm{EE}}}
{\eta_{\mathrm{EE}}^{(\mathrm{ref})}}
\right)
-
\lambda_{\mathrm{QoS}}
\max\!\left(
0,
\gamma_k^{(\mathrm{th})}-\gamma_k
\right)^2
\\
&
+
\lambda_{\mathrm{pwr}}
\left(
1-\frac{P_{\mathrm{agg}}}{P_{\max}}
\right),
\end{aligned}
\label{eq:reward}
\end{equation}
where \(P_{\mathrm{agg}}=\sum_{k=1}^{K}\|\mathbf{v}_k\|_2^2\), \(\eta_{\mathrm{EE}}^{(\mathrm{ref})}\) is a normalization constant, and \(\lambda_{\mathrm{QoS}},\lambda_{\mathrm{pwr}}>0\) are weighting parameters. The first term is a cooperative energy-efficiency reward shared across agents, the second term penalizes QoS violations, and the third term discourages unnecessary power usage. In practice, \(\lambda_{\mathrm{pwr}}\) is selected smaller than \(\lambda_{\mathrm{QoS}}\) so that power reduction does not dominate reliability.

The SINR entering the reward function is computed using the enhanced CDI \(\tilde{\mathbf{g}}_k\) rather than by explicitly solving the worst-case inner minimization in~\eqref{eq:robust_constraint} at each coherence interval. Instead, the learned covariance information is embedded into the local observation vector through \(\mathrm{tr}(\tilde{\mathbf{C}}_k^{(\mathrm{VAE})})/M\), allowing the policy to adapt its beamforming decisions according to CSI reliability while maintaining a fully differentiable online control pipeline.

\subsubsection{Actor--critic updates}

Each agent maintains a deterministic actor
\(\boldsymbol{\mu}_{\boldsymbol{\theta}_k}(\mathbf{o}_k)\) and a centralized critic
\[
Q_{\boldsymbol{\phi}_k}(\mathbf{s},\boldsymbol{\alpha}_1,\ldots,\boldsymbol{\alpha}_K),
\]
which observes all agents' actions during training. The critic is trained by minimizing the temporal-difference loss
\begin{equation}
\mathcal{L}(\boldsymbol{\phi}_k)
=
\mathbb{E}
\left[
\left(
Q_{\boldsymbol{\phi}_k}
(\mathbf{s},\boldsymbol{\alpha})
-
y_k
\right)^2
\right],
\label{eq:critic_loss}
\end{equation}
where
\begin{equation}
y_k
=
r_k
+
\gamma_{\mathrm{RL}}
Q_{\boldsymbol{\phi}_k'}
(
\mathbf{s}',
\boldsymbol{\alpha}_1',
\ldots,
\boldsymbol{\alpha}_K'
).
\end{equation}
Here, \(\gamma_{\mathrm{RL}}\) denotes the RL discount factor to distinguish it from the communication SINR \(\gamma_k\).

The actor is updated using the deterministic policy gradient
\begin{equation}
\nabla_{\boldsymbol{\theta}_k}J
=
\mathbb{E}
\left[
\nabla_{\boldsymbol{\alpha}_k}
Q_{\boldsymbol{\phi}_k}
(\mathbf{s},\boldsymbol{\alpha})
\nabla_{\boldsymbol{\theta}_k}
\boldsymbol{\mu}_{\boldsymbol{\theta}_k}
(\mathbf{o}_k)
\right].
\label{eq:policy_gradient}
\end{equation}
Target networks are updated through soft Polyak averaging \cite{Epasto2020}, and exploration is performed by injecting Gaussian noise into the actor output during training \cite{lowe2017multi}.


Algorithm~\ref{alg:maddpg} summarizes the proposed online beamforming procedure.
\begin{algorithm}[t]
\caption{Physics-informed uncertainty-aware MADDPG beamforming}
\label{alg:maddpg}
\begin{algorithmic}[1]

\REQUIRE
Training replay buffer \(\mathcal{D}\),
trained VAE \((f_{\boldsymbol{\psi}},g_{\boldsymbol{\omega}})\),
actor networks
\(\{\boldsymbol{\mu}_{\boldsymbol{\theta}_k}\}_{k=1}^{K}\),
critic networks
\(\{Q_{\boldsymbol{\phi}_k}\}_{k=1}^{K}\)

\ENSURE
Feasible downlink precoders
\(\{\mathbf{v}_k\}_{k=1}^{K}\)

\STATE \textbf{Offline training phase}

\FOR{each training episode}

    \FOR{each coherence interval \(n\)}

        \FOR{each UE \(k\)}

            \STATE Obtain imperfect CDI
            \(\hat{\mathbf{g}}_k[n]\)

            \STATE Perform VAE inference using
            Algorithm~\ref{alg:vae_csi} to obtain
            \((\tilde{\mathbf{g}}_k,
            \tilde{\mathbf{C}}_k^{(\mathrm{VAE})})\)

            \STATE Construct local observation
            \(\mathbf{o}_k[n]\)
            using~\eqref{eq:local_obs}

            \STATE Generate exploratory action
            \[
            \boldsymbol{\alpha}_k[n]
            =
            \boldsymbol{\mu}_{\boldsymbol{\theta}_k}
            (\mathbf{o}_k[n])
            + \mathbf{z}_k[n]
            \]
            where \(\mathbf{z}_k[n]\) is Gaussian exploration noise

            \STATE Construct raw precoder
            \(\mathbf{v}_k^{(\mathrm{raw})}\)
            using~\eqref{eq:raw_precoder}

        \ENDFOR

        \STATE Project all raw precoders using
        \eqref{eq:power_projection}
        to obtain feasible precoders
        \(\{\mathbf{v}_k[n]\}_{k=1}^{K}\)

        \STATE Compute rewards using~\eqref{eq:reward}

        \STATE Store transition
        \((\mathbf{s}[n],\boldsymbol{\alpha}[n],
        \mathbf{r}[n],\mathbf{s}[n+1])\)
        in replay buffer \(\mathcal{D}\)

        \STATE Sample mini-batch from \(\mathcal{D}\)

        \STATE Update critic networks using
        \eqref{eq:critic_loss}

        \STATE Update actor networks using
        \eqref{eq:policy_gradient}

        \STATE Update target networks via soft Polyak averaging

    \ENDFOR

\ENDFOR

\STATE \textbf{Online deployment phase}

\FOR{each coherence interval \(n\)}

    \FOR{each UE \(k\)}

        \STATE Obtain
        \((\tilde{\mathbf{g}}_k,
        \tilde{\mathbf{C}}_k^{(\mathrm{VAE})})\)
        via Algorithm~\ref{alg:vae_csi}

        \STATE Assemble local observation
        \(\mathbf{o}_k[n]\)

        \STATE Generate action $\boldsymbol{\alpha}_k[n] = \boldsymbol{\mu}_{\boldsymbol{\theta}_k} (\mathbf{o}_k[n])$

        \STATE Construct
        \(\mathbf{v}_k^{(\mathrm{raw})}\)

    \ENDFOR

    \STATE Apply projection~\eqref{eq:power_projection}

\ENDFOR

\RETURN
\(\{\mathbf{v}_k\}_{k=1}^{K}\)

\end{algorithmic}
\end{algorithm}

\section{Simulation Results}
\label{sec:results}

\subsection{Simulation Setup and Baselines}

This section evaluates the proposed framework for robust energy-efficient beamforming in HAPS-enabled massive MIMO systems under imperfect CSI. The evaluation focuses on energy efficiency, SINR robustness, outage performance, and online computational efficiency. All schemes are tested under the same user deployments, channel realizations, feedback impairments, power budget, and QoS thresholds.

Unless otherwise stated, the simulation parameters are listed in Table~\ref{tab:sim_params}. Users are independently and uniformly distributed within the HAPS coverage region and move according to the mobility model described in Section~\ref{sec:system_model}. The channel coefficients are generated from the elevation-dependent Rician model in~\eqref{eq:rician_channel}, including atmospheric attenuation and Doppler-induced temporal correlation. Imperfect BFI is produced using the feedback-delay, quantization, packet-loss, and estimation-noise models in~\eqref{eq:quantized_cdi}--\eqref{eq:packet_loss_cdi}. All reported results are averaged over 500 independent Monte Carlo trials.

The main performance metrics are:
\begin{itemize}
    \item energy efficiency \(\eta_{\mathrm{EE}}\), defined in~\eqref{eq:energy_efficiency};
    \item approximate imperfect-CSI SINR \(\bar{\gamma}_k\), computed using~\eqref{eq:sinr_imperfect_csi};
    \item outage probability, defined as $
    P_{\mathrm{out}}
    =
    \Pr\!\left(\gamma_k < \gamma_k^{(\mathrm{th})}\right)$.
\end{itemize}

The proposed method is compared with the following baselines:
\begin{enumerate}
    \item \textit{Perfect-CSI beamforming}: an ideal benchmark where the HAPS has exact instantaneous CSI and solves the same beamforming objective without CSI uncertainty.
    \item \textit{Imperfect-CSI beamforming}: beamforming based directly on delayed and quantized CDI without uncertainty-aware compensation.
    \item \textit{SDR-based robust beamforming}: a model-based robust optimization baseline using ellipsoidal uncertainty and semidefinite relaxation.
    \item \textit{MADDPG without VAE}: an ablation baseline where the MADDPG controller uses imperfect CDI directly without VAE-based enhancement or learned covariance.
\end{enumerate}

The SDR-based robust beamforming baseline is solved using CVX with the MOSEK solver~\cite{Grant2014CVX}. Since solving the full SDR problem for large arrays becomes computationally prohibitive, the SDR baseline is evaluated at reduced array sizes when needed, while its asymptotic scaling is reported separately in the complexity comparison. Therefore, the reduced-size SDR results are used to illustrate model-based robust optimization behavior, not to claim full-scale real-time feasibility.

The default simulation parameters are summarized in Table~\ref{tab:sim_params}. The carrier frequency, bandwidth, HAPS altitude, feedback delay, quantization resolution, and packet loss probability are chosen to reflect representative mmWave HAPS/NTN operating conditions. The reference Rician factor corresponds to a LoS-dominant channel, while the actual Rician factor varies with elevation according to~\eqref{eq:rician_k_factor}.

\begin{table}[t]
\centering
\caption{Simulation parameters}
\label{tab:sim_params}
\begin{tabular}{|c|c|c|}
\hline
\textbf{Parameter} & \textbf{Symbol} & \textbf{Value} \\ \hline
HAPS altitude & \(H\) & \(20\) km \\ \hline
Coverage radius & \(R_{\mathrm{cov}}\) & \(30\) km \\ \hline
Number of users & \(K\) & \(20\) default, \(10\)--\(100\) varied \\ \hline
User speed & \(v_k\) & \(1\)--\(20\) m/s \\ \hline
Carrier frequency & \(f_c\) & \(28\) GHz \\ \hline
Bandwidth & \(W\) & \(100\) MHz \\ \hline
Noise power & \(\sigma_n^2\) & \(-94\) dBm \\ \hline
UPA size & \(M_x \times M_y\) & \(16 \times 16\) \\ \hline
Total antennas & \(M\) & \(256\) \\ \hline
Maximum transmit power & \(P_{\max}\) & \(40\) W \\ \hline
PA efficiency & \(\eta_{\mathrm{PA}}\) & \(0.38\) \\ \hline
Fixed power & \(P_{\mathrm{fix}}\) & \(120\) W \\ \hline
Feedback delay & \(T_{\mathrm{delay}}\) & \(5\) ms \\ \hline
Quantization bits & \(B_{\mathrm{LoS}}\) & \(6\) \\ \hline
Packet loss probability & \(p_{\mathrm{loss}}\) & \(0.05\) \\ \hline
CSI noise variance & \(\sigma_e^2\) & \(10^{-3}\) \\ \hline
Outage target & \(\epsilon_{\mathrm{out}}\) & \(0.05\) \\ \hline
Reference Rician factor & \(\kappa_{\mathrm{ref}}\) & \(10\) dB \\ \hline
Atmospheric attenuation & \(A(f_c)\) & \(0.12\) dB/km \\ \hline
VAE latent dimension & \(d_z\) & \(16\) \\ \hline
VAE learning rate & -- & \(10^{-3}\) \\ \hline
VAE batch size & -- & \(256\) \\ \hline
RL discount factor & \(\gamma_{\mathrm{RL}}\) & \(0.99\) \\ \hline
Replay buffer size & -- & \(10^6\) \\ \hline
Mini-batch size & -- & \(256\) \\ \hline
Monte Carlo realizations & -- & \(500\) \\ \hline
\end{tabular}
\end{table}

\subsection{Energy-Efficiency and SINR Scaling with User Density}

\begin{figure}[!t]
\centering
\includegraphics[width=\columnwidth]{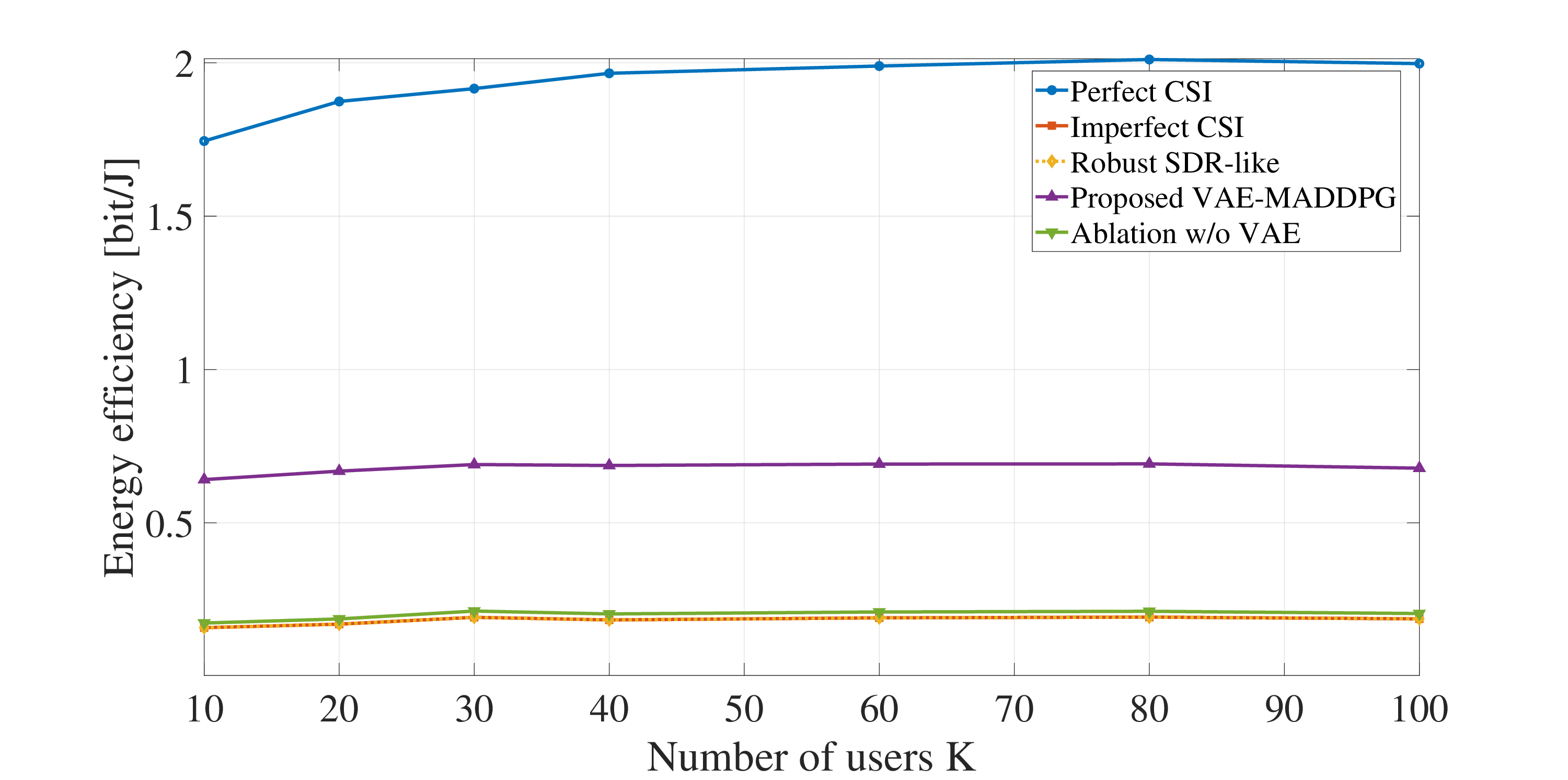}
\caption{Energy efficiency versus the number of users \(K\).}
\label{fig3}
\end{figure}
Fig.~\ref{fig3} shows the energy efficiency as a function of the number of users \(K\). The Perfect-CSI benchmark achieves the highest energy efficiency, as expected, since it has access to exact channel information and can fully exploit the available spatial degrees of freedom. The imperfect-CSI baseline exhibits substantially lower energy efficiency because delayed and quantized CDI leads to inaccurate beamforming, stronger interference leakage, and inefficient power utilization. The SDR-based robust benchmark provides only limited improvement over the imperfect-CSI case, mainly due to its conservative uncertainty treatment.
The proposed VAE--MADDPG framework achieves consistently higher energy efficiency than all practical imperfect-CSI baselines. This gain comes from two effects: the VAE improves the reliability of the CDI representation and provides an uncertainty descriptor, while the MADDPG controller learns a power-aware beamforming policy that avoids overly conservative transmission. The ablation without VAE performs competitively at low user density, but degrades as \(K\) increases, indicating that uncertainty learning becomes increasingly important in dense interference-limited scenarios. Hence, the proposed framework maintains stable energy efficiency as the number of users increases, demonstrating its scalability under imperfect CSI.

\begin{figure}[t]
\centering
\includegraphics[width=\columnwidth]{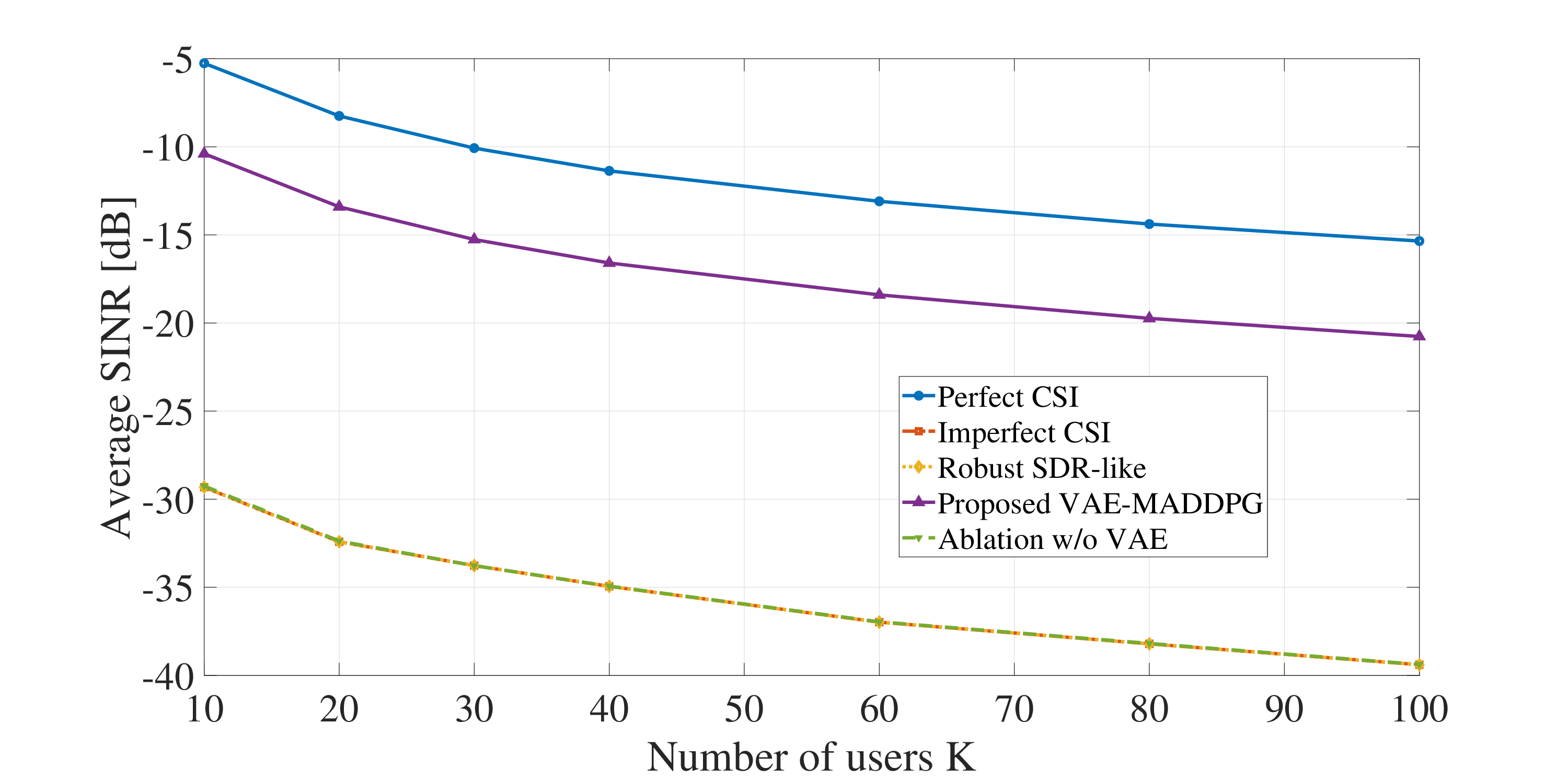}
\caption{Average SINR versus the number of users \(K\) under imperfect CSI.}
\label{fig4}
\end{figure}

Fig.~\ref{fig4} presents the corresponding average SINR versus the number of users. As \(K\) increases, the average SINR decreases for all schemes due to stronger multiuser interference and reduced spatial separation among users. The Perfect-CSI benchmark provides the highest SINR and serves as an upper bound. In contrast, the imperfect-CSI and SDR-based baselines experience significant SINR degradation because inaccurate or overly conservative CSI representations limit their ability to suppress interference.
The proposed VAE--MADDPG framework achieves a clear SINR gain over the practical baselines across the full user-density range. This improvement confirms that the learned CDI enhancement and uncertainty-aware policy help mitigate interference under delayed, quantized, and unreliable feedback. The performance gap between the proposed method and the no-VAE ablation further highlights the importance of the VAE-based uncertainty representation, especially as the network becomes denser. These SINR improvements support the outage and reliability gains reported in the following results.

\subsection{Robustness to Feedback Delay}

\begin{figure}[t]
\centering
\includegraphics[width=\columnwidth]{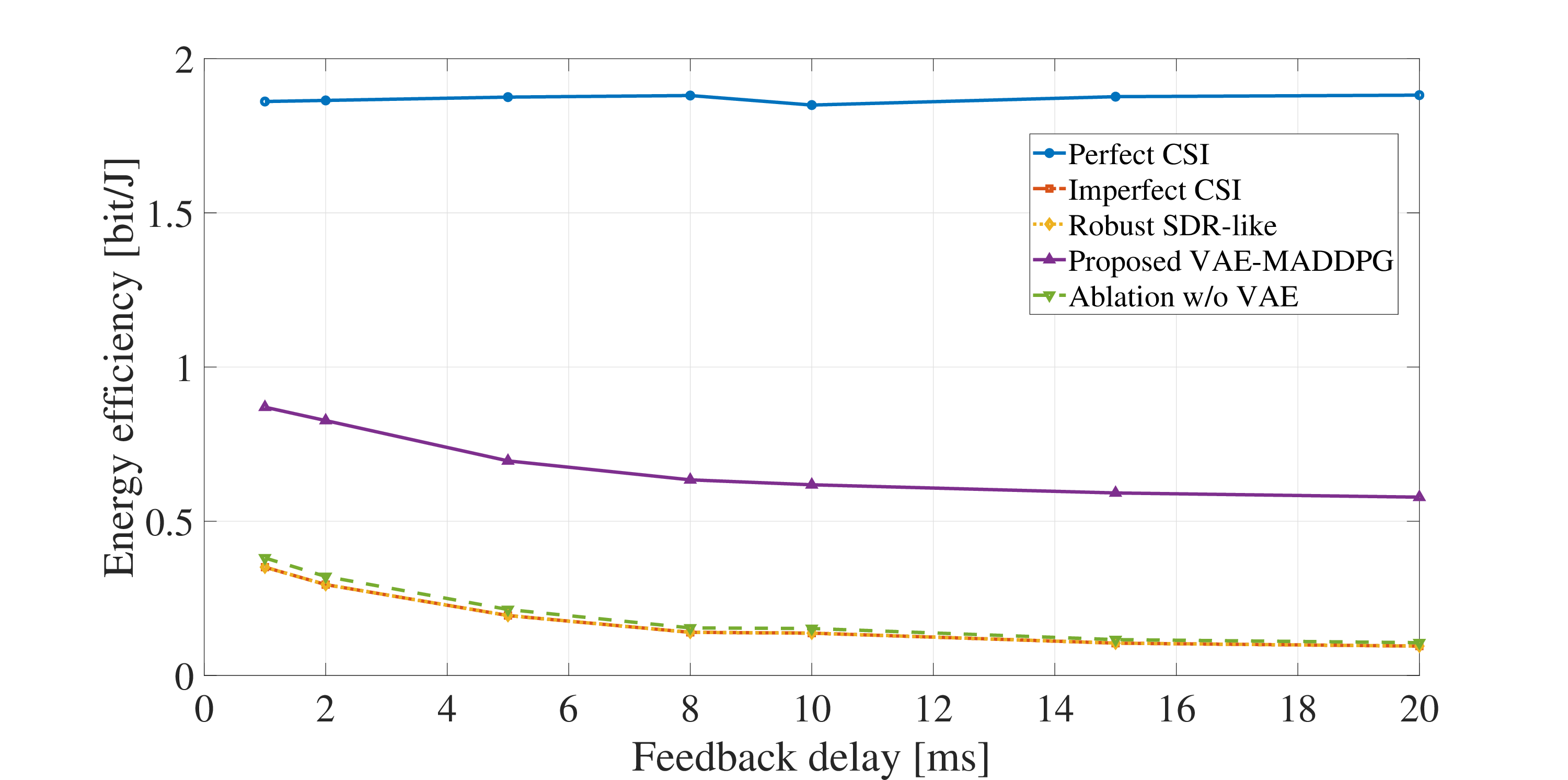}
\caption{Energy efficiency versus feedback delay under imperfect CSI.}
\label{fig5}
\end{figure}

\begin{figure}[t]
\centering
\includegraphics[width=\columnwidth]{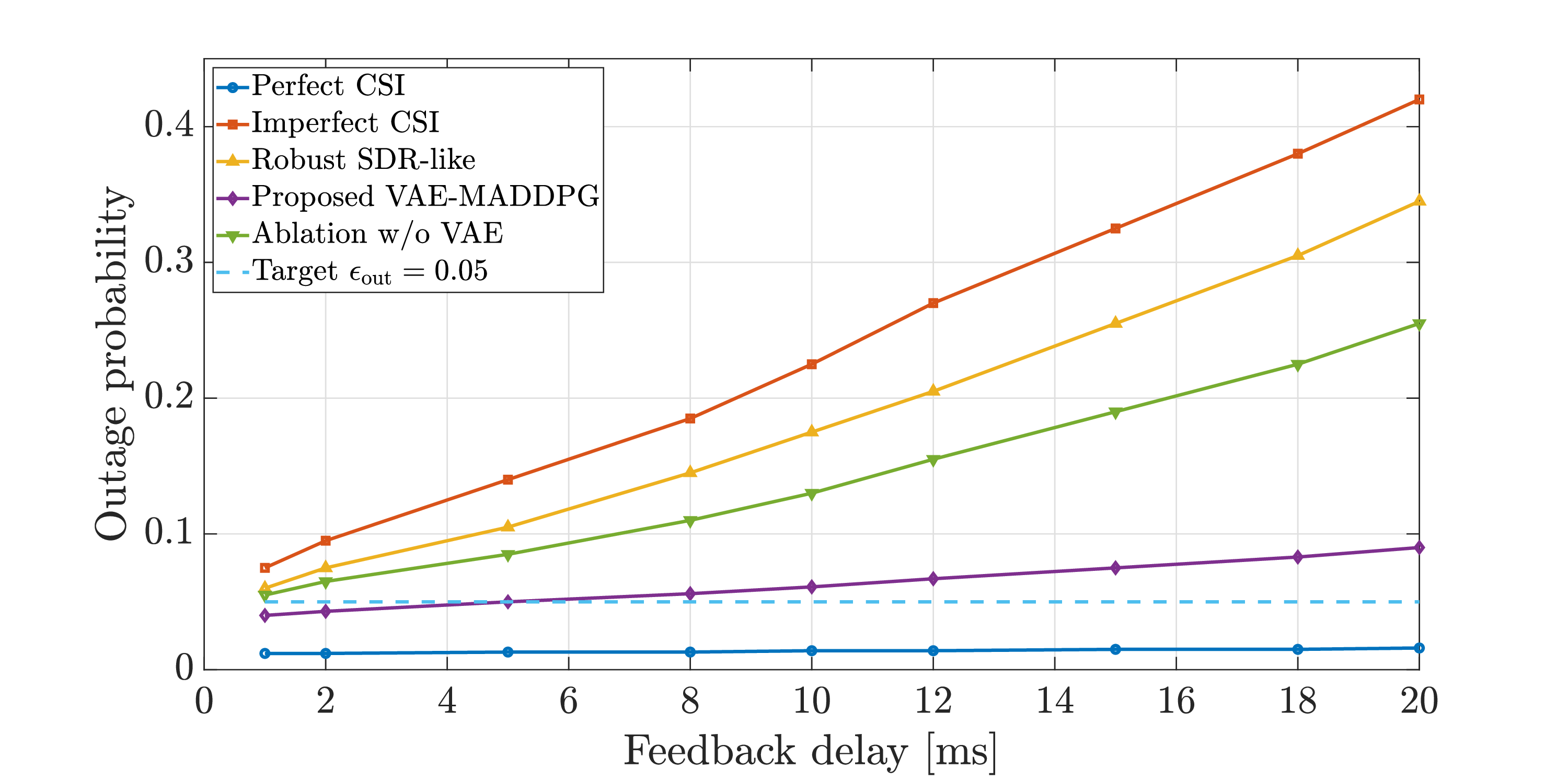}
\caption{Outage probability versus feedback delay under imperfect CSI.}
\label{fig11}
\end{figure}

The robustness of the proposed framework is evaluated under feedback delay, which introduces a temporal mismatch between the estimated and actual channel states. Fig.~\ref{fig5} shows the energy-efficiency degradation as the feedback delay increases, while Fig.~\ref{fig11} presents the corresponding empirical outage probability.

As shown in Fig.~\ref{fig5}, all practical schemes experience energy-efficiency degradation as the feedback delay increases. This behavior is expected because outdated CSI leads to inaccurate beamforming, stronger interference leakage, and less efficient power utilization. The Perfect-CSI benchmark remains nearly constant, since it does not rely on delayed feedback and therefore represents an ideal upper bound. In contrast, the imperfect-CSI baseline and the SDR-based robust baseline both degrade substantially with increasing delay, highlighting the sensitivity of conventional beamforming methods to CSI aging.

The proposed VAE--MADDPG framework exhibits significantly stronger robustness to feedback delay. Its energy efficiency decreases more gradually as the delay increases, indicating that the learned CDI enhancement and uncertainty-aware policy can partially compensate for outdated feedback. The no-VAE ablation consistently underperforms the proposed framework, especially at larger delay values, confirming the importance of the VAE-based uncertainty representation in mitigating channel-aging effects.

Fig.~\ref{fig11} further evaluates the empirical outage probability under increasing feedback delay. The imperfect-CSI baseline exhibits a rapid outage increase because outdated CDI causes beam misalignment and interference leakage. The SDR-based robust baseline improves reliability at small delays but becomes less effective as the temporal mismatch increases. In contrast, the proposed VAE--MADDPG framework maintains the outage probability closer to the prescribed threshold by embedding learned uncertainty information into the policy state. This result supports the probabilistic QoS formulation in~\eqref{eq:prob_qos} and the uncertainty-aware beamforming design in~\eqref{eq:robust_constraint}.

These results show that the proposed framework provides more stable energy-efficiency and reliability performance under delayed feedback compared with conventional imperfect-CSI and SDR-based approaches.

\subsection{Mobility and Doppler Robustness}

\begin{figure}[t]
\centering
\includegraphics[width=\columnwidth]{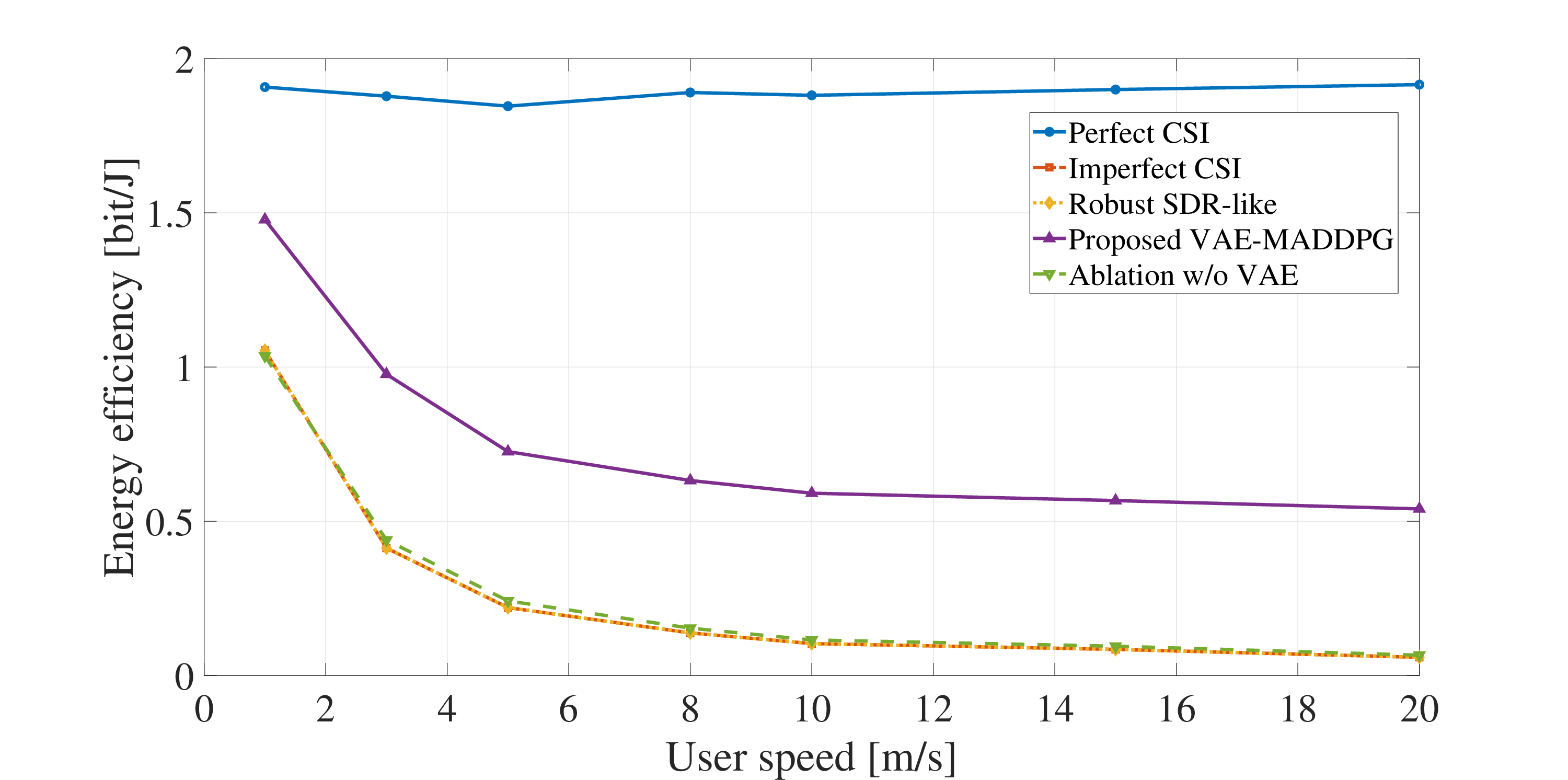}
\caption{Energy efficiency versus user speed under imperfect CSI.}
\label{fig6}
\end{figure}

\begin{figure}[t]
\centering
\includegraphics[width=\columnwidth]{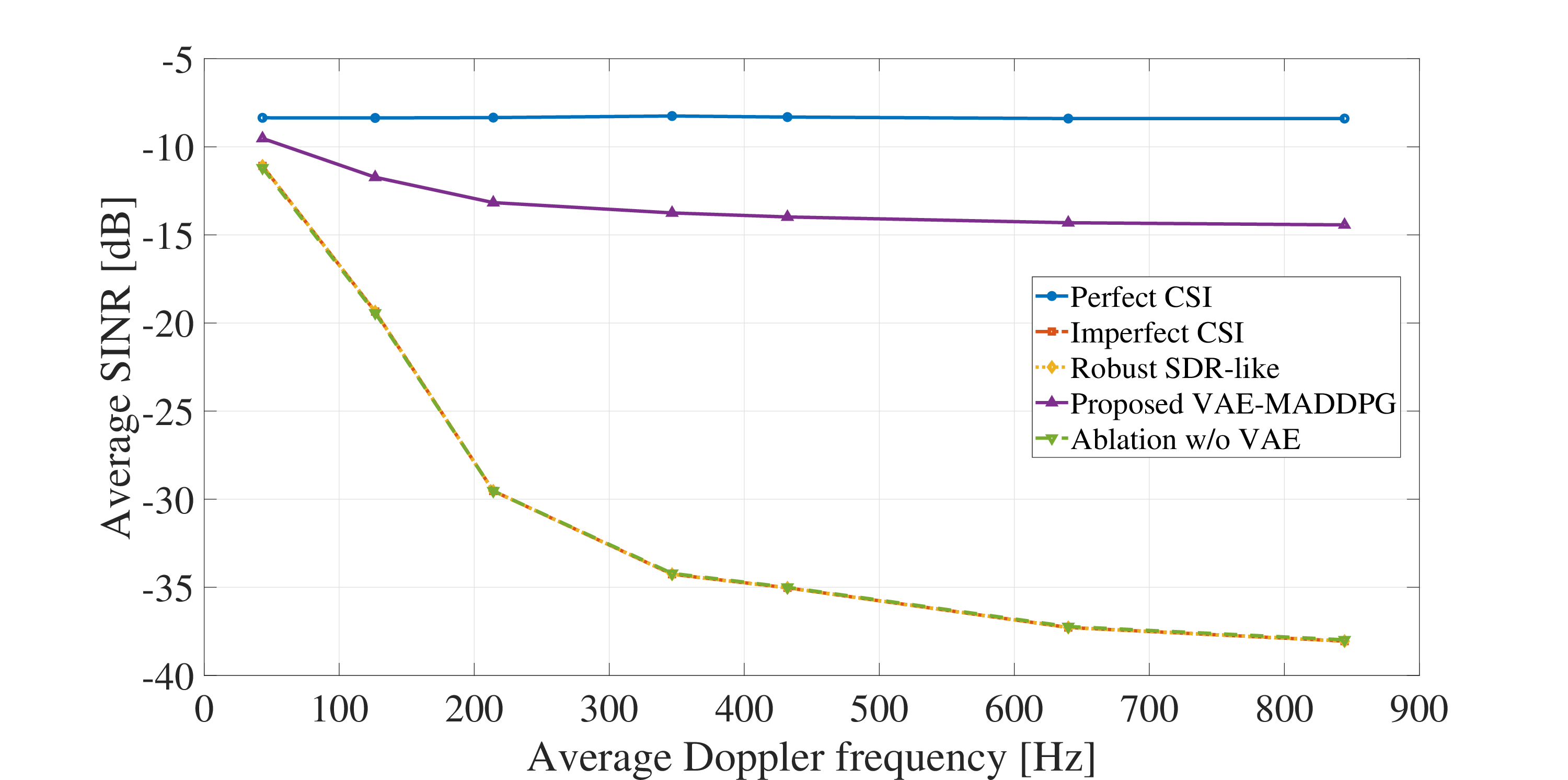}
\caption{Average SINR versus Doppler frequency under imperfect CSI.}
\label{fig7}
\end{figure}

This subsection evaluates the impact of user mobility and Doppler-induced channel variations on the proposed VAE--MADDPG framework. Fig.~\ref{fig6} shows the energy efficiency versus user speed, while Fig.~\ref{fig7} reports the average SINR as a function of Doppler frequency.

As shown in Fig.~\ref{fig6}, all practical schemes experience energy-efficiency degradation as user speed increases. This degradation is caused by faster channel temporal variation, stronger CSI aging, and reduced accuracy of the delayed CDI available at the HAPS. The imperfect-CSI baseline and the SDR-based robust baseline are particularly sensitive to mobility, since their beamforming decisions rely either on outdated CDI or on uncertainty models that become increasingly conservative under rapid channel variation. 
The proposed VAE--MADDPG framework exhibits a more gradual degradation with increasing user speed. This behavior indicates that the learned CDI enhancement and uncertainty-aware policy improve robustness against mobility-induced CSI mismatch. The no-VAE ablation consistently performs below the proposed framework, especially at higher speeds, confirming that VAE-based uncertainty estimation contributes to more reliable beamforming under Doppler-induced channel aging.

Fig.~\ref{fig7} shows the corresponding SINR behavior under increasing Doppler frequency. As the Doppler frequency increases, the imperfect-CSI and SDR-based baselines suffer from stronger SINR degradation due to beam misalignment and residual multiuser interference. In contrast, the proposed framework maintains higher SINR across the considered Doppler range, demonstrating its ability to adapt beamforming decisions according to the learned CSI reliability information. The performance gap between the proposed framework and the no-VAE ablation further confirms that uncertainty-aware CDI enhancement is beneficial in dynamic HAPS channels.

These results show that the proposed framework improves robustness to mobility and Doppler-induced channel aging, maintaining more stable energy-efficiency and SINR performance than conventional imperfect-CSI and SDR-based approaches.

\subsection{Convergence and Online Runtime}

\begin{figure}[t]
\centering
\includegraphics[width=\columnwidth]{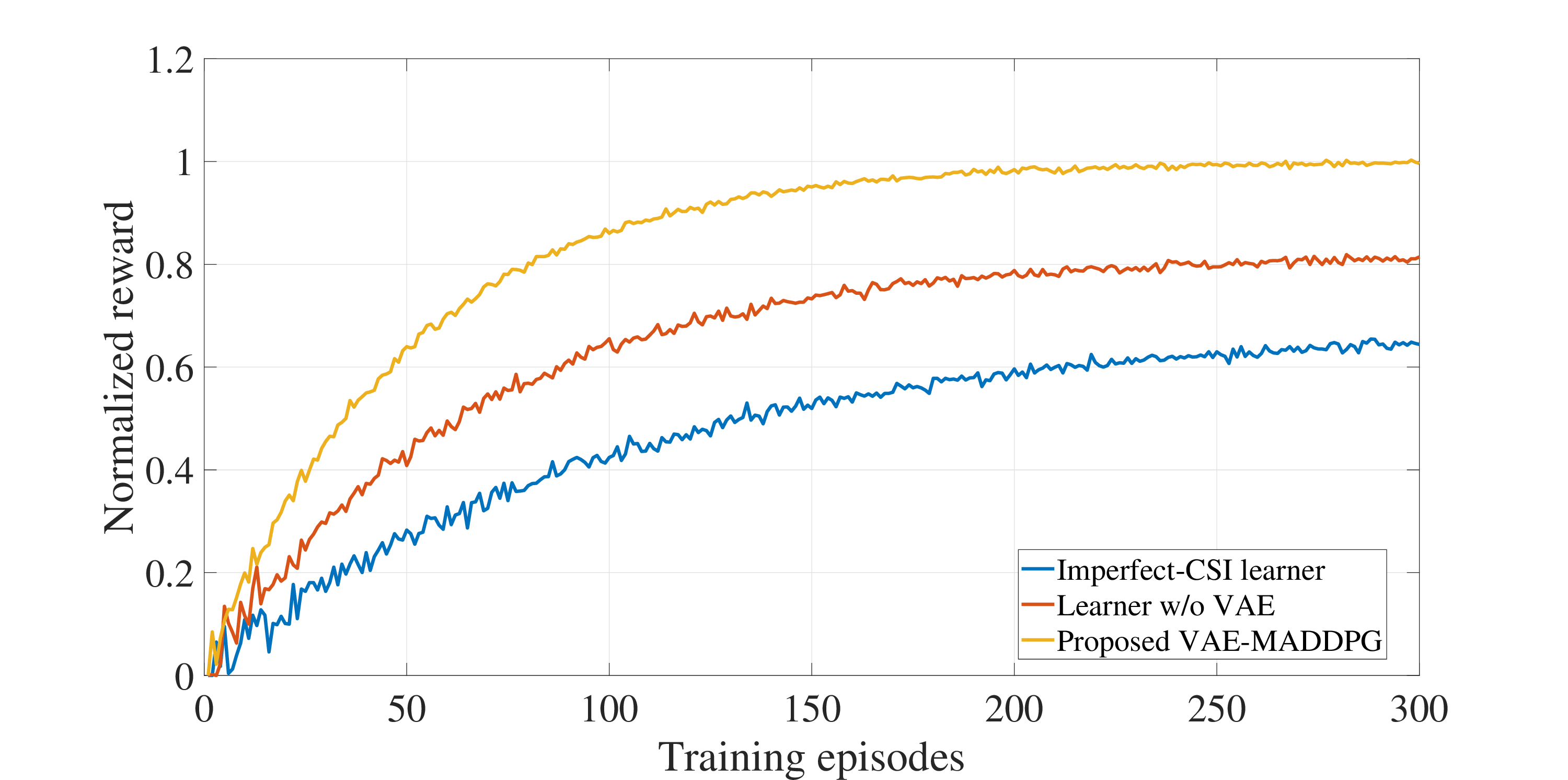}
\caption{Convergence performance in terms of normalized reward versus training episodes.}
\label{fig8}
\end{figure}

\begin{figure}[t]
\centering
\includegraphics[width=\columnwidth]{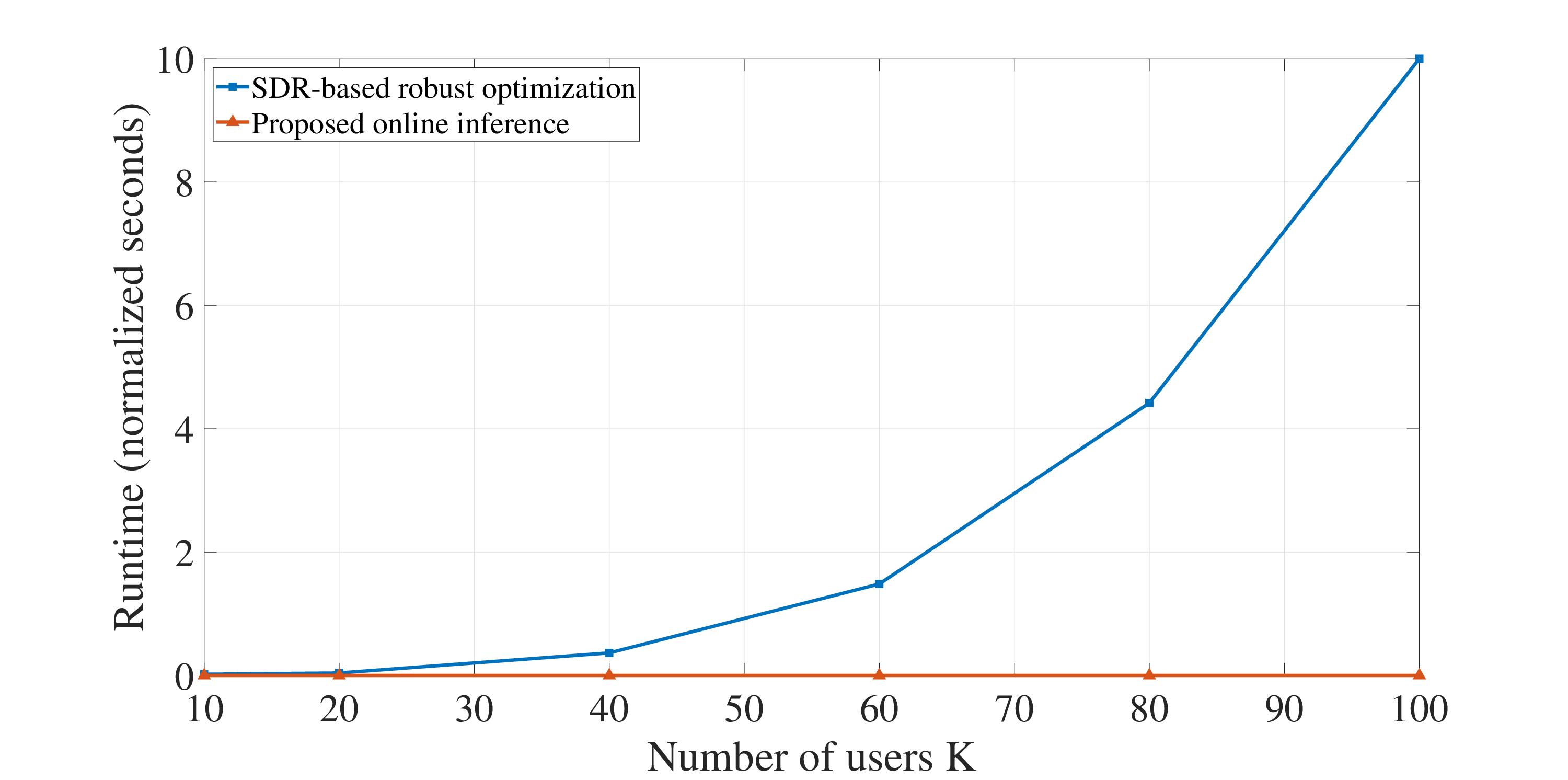}
\caption{Online runtime versus the number of users \(K\).}
\label{fig9}
\end{figure}

The convergence behavior and online computational efficiency of the proposed framework are evaluated in Fig.~\ref{fig8} and Fig.~\ref{fig9}, respectively. Fig.~\ref{fig8} shows the normalized reward versus training episodes, while Fig.~\ref{fig9} compares the online runtime as the number of users increases.

As shown in Fig.~\ref{fig8}, the proposed VAE--MADDPG framework converges faster and reaches a higher normalized reward than the learning-based baselines. The improved convergence behavior is mainly attributed to the VAE-enhanced state representation, which provides the actors with more informative CDI and uncertainty features. In contrast, the no-VAE baseline converges more slowly and stabilizes at a lower reward level, while the imperfect-CSI learner exhibits the weakest convergence due to unreliable state information. These results indicate that incorporating physics-informed uncertainty estimation improves sample efficiency and stabilizes policy learning.

Fig.~\ref{fig9} evaluates the online runtime versus the number of users \(K\). The SDR-based robust optimization baseline exhibits rapidly increasing runtime as \(K\) grows, reflecting the high computational burden of solving a semidefinite program at each coherence interval. In contrast, the proposed framework maintains a much lower online runtime because beamforming decisions are obtained through forward inference of the trained VAE and MADDPG actors, followed by the differentiable projection step. This avoids repeated iterative optimization during deployment and enables scalable online execution for large user densities.

These results confirm that the proposed framework improves learning stability and reduces online computational burden compared with model-based robust optimization, making it suitable for scalable HAPS beamforming under imperfect CSI.
\subsection{Ablation Study and Quantitative Comparison}

\begin{figure}[t]
\centering
\includegraphics[width=\columnwidth]{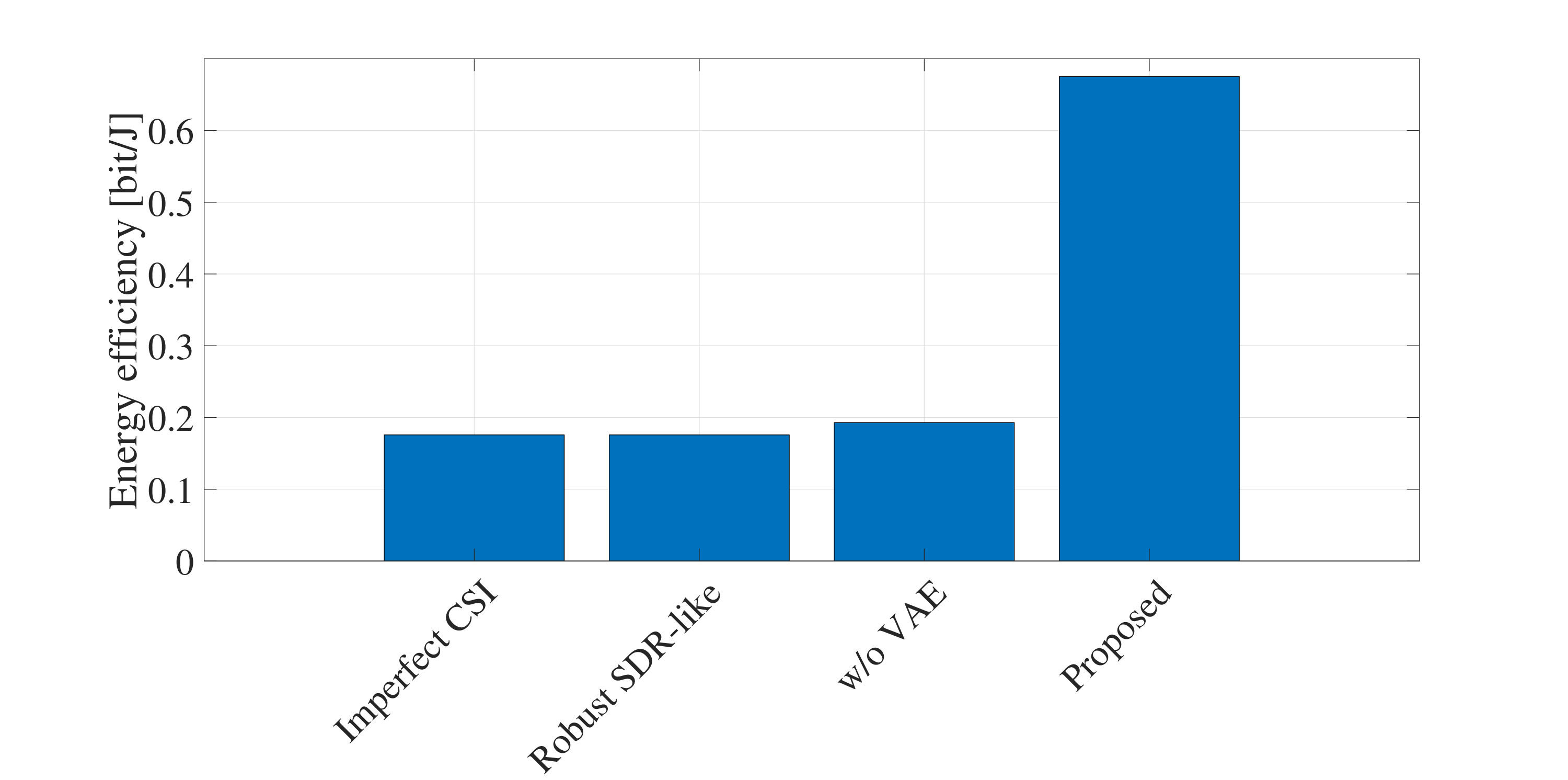}
\caption{Energy efficiency comparison for ablation analysis under imperfect CSI.}
\label{fig10}
\end{figure}

To quantify the contribution of the main components of the proposed framework, Fig.~\ref{fig10} and Table~\ref{tab:ablation_ee} compare the energy efficiency achieved by different schemes under imperfect CSI. 
As shown in Fig.~\ref{fig10}, the imperfect-CSI baseline and the SDR-based robust benchmark achieve comparable energy efficiency. This indicates that model-based robustness alone provides limited energy-efficiency gains under the considered feedback impairments, primarily because conservative uncertainty handling can restrict useful spatial multiplexing and power adaptation. 
The learning-based ablation without VAE provides a modest improvement over the imperfect-CSI baseline, suggesting that the MADDPG controller can partially adapt beamforming decisions even with degraded CSI. However, its gain remains limited because the input channel representation does not  capture the structured uncertainty induced by feedback delay, quantization, packet loss, and Doppler aging.

The proposed VAE--MADDPG framework achieves the highest energy efficiency, reaching \(0.69\) Mbit/J in this setting. As summarized in Table~\ref{tab:ablation_ee}, this corresponds to a \(3.83\times\) gain over the imperfect-CSI baseline. The performance gap between the proposed method and the no-VAE ablation confirms that physics-informed CDI enhancement and learned uncertainty estimation are key contributors to the observed improvement. The ablation results show that combining uncertainty-aware representation learning with multi-agent beamforming control is essential for achieving high energy efficiency under imperfect CSI in HAPS massive MIMO systems.

\begin{table}[t]
\centering
\caption{Energy-efficiency comparison for the ablation study}
\label{tab:ablation_ee}
\renewcommand{\arraystretch}{1.1}
\setlength{\tabcolsep}{3pt}
\begin{tabular}{lccc}
\hline
\textbf{Method} 
& \begin{tabular}{c}\textbf{Energy} \\ \textbf{Efficiency (Mbit/J)}\end{tabular}
& \begin{tabular}{c}\textbf{Gain} \\ \textbf{($\times$)}\end{tabular}
& \begin{tabular}{c}\textbf{Improvement} \\ \textbf{(\%)}\end{tabular} \\
\hline
Imperfect CSI & 0.18 & 1.00$\times$ & 0\% \\
SDR-based robust & 0.18 & 1.00$\times$ & 0\% \\
MADDPG w/o VAE & 0.20 & 1.11$\times$ & 11\% \\
Proposed VAE--MADDPG
& \textbf{0.69} & \textbf{3.83$\times$} & \textbf{283\%} \\
\hline
\end{tabular}
\end{table}

\subsection{Uncertainty Calibration and Outage Validation}
\label{uncertainty}

The preceding results show that the proposed framework improves energy efficiency, SINR, and robustness under imperfect CSI. Since the design relies on the learned ellipsoidal uncertainty set \(\widetilde{\mathcal{U}}_k\) in~\eqref{eq:learned_set}, it is also necessary to assess whether this set provides a calibrated representation of the CDI error. In particular, the approximate probabilistic QoS interpretation in~\eqref{eq:prob_qos}--\eqref{eq:robust_constraint} depends on how accurately the learned uncertainty region captures the true CDI perturbations.

Table~\ref{tab:calibration} reports the empirical coverage probability of \(\widetilde{\mathcal{U}}_k\) for different target confidence levels. The empirical coverage is computed over independent channel realizations as the fraction of true CDI samples that fall inside the learned ellipsoid. As shown in the table, the learned uncertainty set closely follows the desired confidence levels. For example, at the 95\% target confidence level, the empirical coverage is 94.2\%, indicating that the VAE-calibrated covariance provides a reliable approximation of the CDI uncertainty.  
The small mismatch between target and empirical coverage is expected due to the tangent-space Gaussian approximation, finite-sample effects, and residual covariance-estimation errors in the VAE output. Nevertheless, the empirical coverage remains close to the target confidence level across all tested cases, confirming that the learned uncertainty set provides a meaningful reliability measure rather than only a numerical artifact.

\begin{table}[t]
\centering
\caption{Calibration of learned ellipsoidal CDI uncertainty sets}
\label{tab:calibration}
\begin{tabular}{c c c}
\hline
Target confidence & Empirical coverage & Empirical outage \\
\hline
90\% & 88.9\% & 10.8\% \\
95\% & 94.2\% & 5.6\% \\
99\% & 97.8\% & 2.4\% \\
\hline
\end{tabular}
\end{table}

The last column of Table~\ref{tab:calibration} reports the corresponding empirical outage probability. The outage probability decreases as the target confidence level increases, which is consistent with the larger uncertainty regions used for more conservative beamforming. At the 95\% confidence level, the empirical outage is 5.6\%, close to the prescribed target \(\epsilon_{\mathrm{out}}=0.05\). This supports the approximate probabilistic QoS formulation and confirms the practical relevance of the learned uncertainty set.
Combined with the outage-delay behavior in Fig.~\ref{fig11}, these calibration results show that the proposed physics-informed VAE not only enhances the CDI estimate but also provides uncertainty information that is useful for reliability-aware beamforming.

\section{Conclusion}

This paper developed a physics-informed uncertainty-aware beamforming framework for HAPS-enabled massive MIMO systems under imperfect CSI. A geometry-aware channel and feedback-impairment model was first established by incorporating elevation-dependent Rician fading, Doppler-induced channel aging, finite-rate feedback quantization, packet loss, and estimation noise. These impairments were mapped into tangent-space ellipsoidal CDI uncertainty sets, enabling an approximate probabilistic QoS interpretation. To refine imperfect CDI and quantify uncertainty, a physics-informed VAE was designed using the LoS-dominant steering structure, unit-sphere CDI projection, and covariance propagation. The learned uncertainty representation was then embedded into a MADDPG-based beamforming framework with CTDE uncertainty-aware local observations, and differentiable power projection. Simulation results showed that the proposed framework improves energy efficiency, SINR robustness, outage reliability, convergence behavior, and online runtime compared with imperfect-CSI, SDR-based, and no-VAE baselines. These results indicate that  modeling CSI uncertainty and embedding it into the beamforming policy is essential for practical HAPS massive MIMO operation under delayed, quantized, and unreliable feedback.

\bibliographystyle{IEEEtran}
\bibliography{IEEEabrv,References}
\end{document}